\documentclass[conference]{IEEEtran}
\pdfoutput=1
\usepackage[english]{babel}

\usepackage[colorinlistoftodos]{todonotes}

\usepackage{booktabs,caption}
\usepackage[flushleft]{threeparttable}
\usepackage{lstlinebgrd}
\usepackage{booktabs}
\usepackage{tabularx}
\usepackage{longtable,color,psfrag,dsfont,supertabular,bm,units}
\usepackage{amssymb,amsmath}
\usepackage{graphicx} 
\usepackage{algorithmic}
\usepackage{multirow}
\usepackage{multicol}
\usepackage{subfigure}
\usepackage{fullpage}
\usepackage{lscape}
\usepackage{afterpage}
\usepackage{tablefootnote}
\usepackage{url}
\usepackage{xcolor, colortbl}
\usepackage{comment}
\usepackage{hyperref}
\PassOptionsToPackage{hyphens}{url}\usepackage{hyperref}
 \usepackage[ruled,vlined,linesnumbered]{algorithm2e}
\usepackage{amssymb,amsmath}
\usepackage[hang,flushmargin]{footmisc}
\usepackage{lipsum}
\usepackage[hyphenbreaks]{breakurl}
\usepackage[hyphens]{url}

\usepackage{color}
\definecolor{light-gray}{gray}{0.90}
 \makeatletter
\newcommand{\algorithmfootnote}[2][\footnotesize]{%
 \let\old@algocf@finish\@algocf@finish
 \def\@algocf@finish{\old@algocf@finish
    \leavevmode\rlap{\begin{minipage}{\linewidth}
    #1#2
    \end{minipage}}%
  }%
}
\makeatother
\usepackage{geometry}
\iftrue
\newcommand{\leo}[1]{{\color{brown}{\bf LB comments:} #1}}
\else
\newcommand{\leo}[1]{}
\fi
\iftrue

\else
\newcommand{\leo}[1]{}
\fi

\makeatletter 
\newcommand\semiHuge{\@setfontsize\semiHuge{22.72}{27.38}}
\makeatother 

\usepackage{listings}
\usepackage{listings}
\usepackage{color}
\DeclareCaptionFont{black}{\color{black}}
\newcommand{\bcircle}{
\begin{tikzpicture}
\filldraw[fill=black,draw=black] circle (2pt);
\end{tikzpicture}
}

\newcommand{\wcircle}{
\begin{tikzpicture}
\filldraw[fill=white,draw=black] circle (2pt);
\end{tikzpicture}
}
 
\begin{document}
\title{IoTDots: A Digital Forensics Framework for Smart Environments}

\author{\IEEEauthorblockN{Leonardo Babun, Amit K. Sikder, Abbas Acar, and A. Selcuk Uluagac}
\IEEEauthorblockA{Cyber-Physical Systems Security Lab \\ Department of Electrical \& Computer Engineering,
Florida International University\\
10555 West Flagler St. Miami, FL 33174\\
Email:  \{lbabu002, asikd003, aacar001, suluagac\}@fiu.edu}
}

\maketitle

\begin{abstract}
IoT devices and sensors have been utilized in a cooperative manner to enable the concept of a smart environment. In these smart settings, abundant data is generated as a result of the interactions between devices and users' day-to-day activities. Such data contain valuable forensic information about events and actions occurring inside the smart environment and, if analyzed, may help hold those violating security policies accountable. Nonetheless, current smart app programming platforms do not provide any digital forensics capability to identify, trace, store, and analyze the IoT data. To overcome this limitation, we introduce IoTDots, a novel digital forensic framework for a smart environment such as smart homes and smart offices. IoTDots has two main components: IoTDots-Modifier and IoTDots-Analyzer. At compile time, IoTDots-Modifier performs the source code analysis of smart apps, detects forensically-relevant information, and automatically insert tracing logs. Then, at runtime, the logs are stored into a IoTDots database. Later, in the event of a forensic investigation, the IoTDots-Analyzer applies data processing and machine learning techniques to extract valuable and usable forensic information from the devices' activity. 
In order to test the performance of IoTDots, we tested IoTDots in a realistic smart office environment with a total of 22 devices and sensors. Also, we considered 10 different cases of forensic activities and behaviors from users, apps, and devices. The evaluation results show that IoTDots can achieve, on average, over 98\% of accuracy on detecting user activities and over 96\% accuracy on detecting the behavior of users, devices, and apps in a smart environment. Finally, IoTDots performance yields no overhead to the smart devices and very minimal overhead to the cloud server. To the best of our knowledge, IoTDots is the first lightweight forensic solution for IoT devices that combines the collection of the forensically-relevant data from a smart environment and the analysis of such data using data processing and machine learning techniques for forensic purposes. Finally, we have made the IoTDots-Modifier available online for the community.
\end{abstract}

\section{Introduction}
The Internet of Things (IoT) has quickly evolved as the network of Internet-enabled physical devices. The IoT devices communicate with each other and interact with the users' day-to-day activities through sensors. These capabilities have also enabled the concept of a \textit{smart environment}. Such an environment improves the quality of the life of the people while handling a new set of data previously untapped. For instance, smart offices can be equipped with a set of state-of-the-art gadgets like smart lights, smart surveillance cameras, smart smoke detectors, presence sensors, and smart thermostats, all controlled by a central hub. In such a setup, the presence sensor can detect users and trigger the cameras to start recording while turning on the lights. At the same time, the smart thermostat can become active and start lowering the room temperature. In this example, the smart environment has control and gets feedback from the users' activities to change the general state of the surroundings based on (1) what the users do, (2) the smart environment setup policies, and (3) the state of the devices. 

Indeed, the interactions between devices and users in a smart environment generate data with tremendous forensic value. For instance, in the smart office setup previously described, the changes in the presence sensor's state may indicate the existence of individuals inside the smart office during unauthorized hours. Further, the smart environment data can give insights about the exact location of abrupt spikes in temperature values or the presence of smoke happening right before a fire incident. 

Nonetheless, current IoT solutions do not provide any means for forensic analysis. In general, the limitation of available computing resources in the majority of the smart devices and the unique cloud-based architecture of IoT makes it very challenging to store data inside the devices for forensic purposes. Indeed, the most popular IoT programming platforms (e.g., Samsung SmartThings, openHab, etc.) do not provide any means to have access and indefinitely store data in the cloud. Previous works have used the idea of logging to save data from smart apps and devices for further analysis of data provenance. For instance, in \cite{fear}, the authors propose an instrumental platform-centric approach that logs activities from smart apps for data provenance. In that work, however, the analysis considered the relationship between devices, users, and smart apps while assuming device integrity. This approach, while valid and useful, overlook the problem of tampered devices \cite{icc} that can cause the abnormal behavior of the smart environment. Without this analysis, the IoT forensic investigation would ignore additional threats coming from the compromised devices \cite{thingspeak, camera}. These threats can modify the state of the smart environment by poisoning the forensic data or by leaking sensitive information that attackers can use to bypass the security policies of the smart environment and gain access, for instance, to restricted areas.  
In this paper, we introduce IoTDots, a novel digital forensic framework for smart environments. IoTDots has two main components: IoTDots-Modifier (ITM) and IoTDots-Analyzer (ITA). The ITM analyzes smart applications looking for forensically-relevant information inside the apps. Then, the smart apps are modified by inserting specific logs that will send the forensic data to the IoTDots Logs Database (ITLD) at runtime. Later, in a case of a forensic investigation, the ITA applies data processing and machine learning techniques on the ITLD data to comprehensively learn the state of the smart environment and the users' behavior at the time of interest of the forensic analysis. IoTDots considers the events and actions inferred from the ITLD data and the security policies defined for the smart environment to detect potential security violations from users, devices, or smart apps. 

Specifically, IoTDots will look for users performing activities that potentially violate the security policies inside a smart environment. Also, IoTDots will use the relationship between devices' states to detect users trying to remove, disable, or tamper IoT devices and/or malicious apps modifying the data extracted from the smart environment. 

We evaluate IoTDots in a smart environment containing different types of devices including smart lights, smart thermostats, motion sensors, etc. Experimental results demonstrate that IoTDots performs very well on detecting, storing, and processing forensically-relevant data from the smart environment. Specifically, for activity detection, our framework achieves up to 98\% accuracy for both time-dependent\footnote{Activities performed only over certain specific time frame \texttt{t}.} and time-independent\footnote{Activities performed freely without considering any specific time frame.} activities. On the other hand, for the case of behavior detection, our framework achieves 96\% accuracy. Additionally, a detailed performance analysis shows that no overhead is imposed to the smart devices and minimal overhead is imposed to the use of physical memory and latency on the cloud-based servers as a result of utilizing IoTDots. 

\noindent\textbf{Summary of Contributions}: The contribution of this work are as follows:

\begin{itemize}
    \item We propose IoTDots, a novel digital forensic framework for smart settings. Our framework automatically analyzes and modifies smart apps to detect and store forensically-relevant data from smart devices, apps, and users. Then, in the case of a forensic investigation, IoTDots applies data processing and machine learning techniques to detect valuable forensic evidence from smart devices, apps, and users' activities and behavior.
    \item We made the IoTDots-Modifier freely available online at 
    so IoTDots users can use our automated system to perform source code analysis and modification of their smart apps to enable IoTDots. 
    \item We used a threat model that considered users performing activities that violate the security policies inside a smart environment. Also, our threat model protects device and data integrity by considering the cases of malicious users tampering smart devices and apps poisoning the smart environment data or misreporting device states.
    \item We evaluated IoTDots in a realistic smart office setup that contains 10 different types of smart devices including smart lights, smart thermostats, motion sensors, etc. and 22 devices in total.
    \item Our results demonstrate that IoTDots achieves very high accuracy on revealing forensic-relevant activities and behavior inside the smart environment.
    \item Finally, since IoTDots focuses on forensic-relevant data only, its performance does not represent additional overhead in terms of computing resources to the smart devices and minimal overhead to the cloud-based servers. 
\end{itemize}

\noindent\textbf{Organization}: The rest of this paper is organized as follows. In Section~\ref{sec:background}, we present the background information. Then, a use case and the threat model are described in Section \ref{sec:threatmodel}. Section \ref{sec:architecture} details the architecture of IoTDots and Section \ref{sec:analysis} describes the data processing and machine learning techniques utilized by the proposed forensic framework. Then, the evaluation results are analyzed in Section \ref{sec:evaluation}, followed by the benefits, challenges, limitations of IoTDots and future work in Section \ref{sec:benefits}. Finally, the related work in Section \ref{sec:related} and the conclusions in Section \ref{sec:conclusion} complete the paper.

\section{Background} \label{sec:background}

\noindent\textbf{Digital Forensics}: The term \textit{digital forensics} refers to the investigation and the uncovering of any kind of evidence from electronic devices containing digital data. The most basic application of digital forensics investigations is to either support or oppose the hypothesis proposed in a case before a court. Frequently, digital forensics requires the analysis of physical evidence directly related to people's presence in the event scene such as an audio or video file. In other cases, a more deep investigation of more technical evidence such as network data or mobile device forensics may be required. Complex cases may require the combination of both types of analysis. 

Although there are different approaches, a standard digital forensics process can be completely described with five different stages~\cite{hegarty2014digital, phases}: (1) Identification, (2) Preservation, (3) Recovery, (4) Analysis, and (5) Presentation. Depending on the type of device analyzed and the source of the evidence, the methodology used and the type of evidence extracted may differ. For example, while a mobile device is a good source of information for tracking the location of a person, a computer may have the capability to provide more detailed and varied intelligence. 

\noindent\textbf{IoT Forensics}: With the growth of IoT, the digital forensics is not anymore limited to storage devices like USB drives, computers, or smartphones. Now, the data from devices like a smart lock,  smartwatch, or motion sensors can also be used for forensic purposes \cite{forensiciot, forensiciot2}. The heterogeneous data that these devices can provide based on their interaction with users' day-to-day activities are very valuable. For example, a smart lock can reveal when and who entered the building or a smart speaker may reveal the exact location of its user during the time of an incident. However, the use of data from smart devices is not a straightforward exercise due to some unique challenges present in the IoT:

\noindent\textbf{Variety of network protocols}: Although there are some popular wireless communication protocols used for communication purposes such as WiFi or ZigBee, there exists no common standard for all smart devices \cite{magazine}. This situation can generate inconsistencies in the smart setup topologies. For instance, in some cases of ZigBee-enabled devices, a smart hub is required to connect the devices to the internet while in WiFi-enabled devices, a direct connection to a gateway is possible. As a consequence of this diversity, it is difficult to identify the source of the evidence data in most cases.
\begin{itemize}
\item\textit{Limited computing resources}: Most of the smart devices are very limited in terms of power, computational resources, and storage capacity \cite{icc}. These restrictions make it harder to store and process the data inside the devices for forensic purposes. To solve this limitation, an alternative would be to transmit and store all the data to a remote server. 
\item\textit{High divergence and diversity in smart data}: In traditional digital forensics investigations, the data is normally collected from specific devices of interest. However, in IoT forensics, the data can be collected from a diverse set of devices \cite{variety}. Different data sources may have different impact and meaning during a digital forensic investigation. Therefore, it is required to have a reliable mechanism that can successfully handle the data collected from different types of smart devices.
\end{itemize}

\section{Threat Model and Assumptions} \label{sec:threatmodel}
In this section, we describe the assumptions focusing on a use case where IoTDots can be utilized to perform forensic investigations. Moreover, we explain the details of the threat model assumed to classify forensically-valuable anomalous user activities and malicious behavior from users and smart apps in a smart environment. 

\subsection{Assumptions}
This work assumes that there exists an office $O$. The office has several devices integrated to create a fully equipped smart environment. The topology of the smart environment in $O$ includes devices like smart thermostats, smart locks, smart lights, presence sensors, security cameras, and smart smoke detectors. We also assume that the general manager Bob is the only person in $O$ with administrative rights to handle the apps that control the smart devices. Such apps have been previously modified by the IoTDots framework and, by policy, they are the only ones authorized to be used to manage the devices inside $O$. Finally, the security policies of $O$ prohibit the presence of any person between 8:00 pm to 7:00 am from Monday through Friday and anytime during the weekends. At some point, a fire incident inside $O$ has caused the loss of sensitive information along with important economic consequences. The fire department has determined that the fire was generated during the night time and a forensic investigation is requested. 

We introduce IoTDots as a novel framework that can utilize the logs from the IoT-modified apps to perform the forensic analysis of the event in $O$. Indeed, IoTDots can be used in conjunction with traditional forensic analysis tools and techniques to hold the responsible person, smart app, or device (if any) accountable if a case of negligence or deliberate violation of security policies is detected. By using IoTDots, several forensic-related questions can be answered: (1) What was happening inside $O$ right before the fire incident had occurred (e.g., right before the smart smoke detector started sensing the smoke presence)?; (2) Was anyone inside the room (e.g., presence sensor state changed)?; (3) Was the door opened/closed anytime before the fire incident (e.g., smart locker state changed)?. Further, the proposed framework would be able to evaluate the different states of the smart environment inside $O$ and match them with the security policies in place to detect any (intentional or involuntary) violation of security policies (e.g., Bob accessing the office at night time) or malicious activities from users (e.g., Bob tampering the security camera to avoid video recording) or smart apps (e.g., apps containing malicious functions to modify the values from the presence sensor possibly by a remote attacker). 

\begin{table*}[t]
\centering
\small
\begin{tabular}{cccc}
\toprule
\textbf{\begin{tabular}[c]{@{}c@{}}Threat\end{tabular}} & \textbf{\begin{tabular}[c]{@{}c@{}}Time-dependency\end{tabular}} & \textbf{\begin{tabular}[c]{@{}c@{}}Attack Method\end{tabular}} & \textbf{\begin{tabular}[c]{@{}c@{}}Specific Attack Examples\end{tabular}} \\ \hline
\midrule
\rowcolor{light-gray}
Activity-1    & Time-independent  & Tampered device   & \begin{tabular}[x]{@{}c@{}}Bob changes the orientation of the presence sensor\\ to fit in a new equipment\end{tabular} \\ 
Activity-2    & Time-independent  & Authorized user   & \begin{tabular}[x]{@{}c@{}}Bob manually lowers the temperature of smart thermostat from\\home the night before of a big meeting with the stakeholders\end{tabular} \\ 
\rowcolor{light-gray}
Activity-3    & Time-independent  & Authorized user   & Bob is inside the office $O$ at 8:45 pm\\ 
Activity-4    & Time-dependent    & Authorized/unauthorized user  & Bob is getting into the office at 8:45 pm \\
\rowcolor{light-gray}
Activity-5    & Time-dependent    & Authorized/unauthorized user  & Bob is using the secure pin to unlock the smart lock  \\
Behavior-1 & Time-independent  & Tampered device  & \begin{tabular}[x]{@{}c@{}}Bob disables the smart camera to stop recording while he is \\performing unauthorized activities inside the office $O$\end{tabular}         \\
\rowcolor{light-gray}
Behavior-2 & Time-independent  & Unauthorized user/attacker    & \begin{tabular}[x]{@{}c@{}}Alice gets access to the office $O$ using the smart lock pin \\that she obtained through a malicious app that leaks information\end{tabular}        \\
Behavior-3 & Time-independent  & Malicious app & The presence sensor app reports inverted states to IoTDots  \\
\rowcolor{light-gray}
Behavior-4 & Time-independent  & Malicious app & \begin{tabular}[x]{@{}c@{}}The smoke detector app triggers the smart windows \\and locks open by reporting false user presence\end{tabular}       \\
Behavior-5 & Time-independent  & Malicious app & \begin{tabular}[x]{@{}c@{}}The smart light app disable the compromised smart camera \\by creating a specific light on/off pattern\end{tabular}         \\
\bottomrule
\end{tabular}
\caption{Summary of the threat model and forensically-valuable activities and behavior considered for IoTDots.}
\label{tab:threatModel}
\end{table*}

\subsection{Threat and Forensic Model}
IoTDots analyzes data from a smart environment to provide answers during the course of a forensic investigation. These answers should provide enough information to either exonerate innocent people or hold the responsible person accountable for his/her actions. To achieve that, the framework needs to be able to successfully detect and characterize any forensically-valuable information, including regular user activities, anomalous user activities, and malicious behavior from users, smart apps, and devices. Within the context of this work, we define these operations as follows: 
\begin{itemize}
\item \textit{Regular User Activities:} Any action performed by the authorized users inside the smart environment that does not violate the security policies in place. 
 \item \textit{Anomalous User Activities:} Any careless and unintentional action performed by an authorized user inside the smart environment that is considered a violation of the security policies in place. 
 \item \textit{Malicious Behavior:} Any intentional action performed by authorized users, non-authorized users, smart apps, and/or smart devices that clearly violates the security policies and is considered a threat for other users and/or the state of the smart environment. 
\end{itemize}
Hence, for IoTDots, we consider two different sets of forensically-valuable threats: (1) anomalous user activities and (2) malicious behavior from users or smart apps. 

To evaluate IoTDots in detecting anomalous user activities, we consider the following time-based activities:
\begin{itemize}
\item\textit{Time-independent Activity 1}: An authorized user controlling a smart device in an anomalous manner inside the smart environment at any time.
\item\textit{Time-independent Activity 2}: An authorized user controlling a smart device in an anomalous manner from outside of the smart environment at any time.
\item\textit{Time-independent Activity 3}: An authorized user having an unauthorized presence in a specific area within the smart environment at any time.
\item\textit{Time-dependent Activity 4}: An authorized/unauthorized user moving inside the smart environment at an unusual time.
\item\textit{Time-dependent Activity 5}: An authorized/unauthorized user accessing the smart environment at an unusual time. 
\end{itemize} 

Since IoTDots considers the security policies of a smart environment to perform its analysis, we categorize user activities as \textit{time-independent} and \textit{time-depended}. In the first group, we include all the user actions whose execution time is irrelevant to IoTDots as it is not regulated as part of the security policies. From the previous list, Activity-1, -2, and -3 can be considered as time-independent actions as they do not depend on time. On the other hand, the second group gathers those actions whose execution time is strictly regulated by the security policies of the smart environment. In this category, we include Activity-4 and -5 because they can only be considered forensically-valuable during a certain period of time (e.g., the presence of an authorized user in $O$ is only considered a violation between 8:00 pm and 7:00 am).
 
Further, we also consider the following forensic behaviors from users or smart apps as part of the threat model for IoTDots.
\begin{itemize}
\item\textit{Behavior 1}: An authorized user can try to tamper, destroy, or remove devices to prevent that the IoTDots logs can be acquired and sent to the ITLD. 
\item\textit{Behavior 2}:  An attacker uses a malicious app to get unauthorized physical or logical access to the smart environment. This threat represents a case of impersonation attack.
\item\textit{Behavior 3}: A malicious app reporting incorrect states (e.g., fake logs) to IoTDots to cover malicious activities. This threat represents a case of false data injection attack.
\item\textit{Behavior 4}: A malicious app installed in the smart environment changes the normal behavior of other smart home devices and applications by enforcing incorrect device states. This threat represents a case of denial-of-service attack.
\item\textit{Behavior 5}: A malicious app triggers malicious actions in other devices and apps by executing a specific activity pattern. This threat represents a case of a side channel attack.
\end{itemize}

Finally, Table \ref{tab:threatModel} summarizes the threat model specifying attack examples, the time-dependency categorization for every action/behavior, and the methods of the attacks.

\section{IoTDots Architecture}\label{sec:architecture}
In this section, we present the general architecture of IoTDots. Also, we detail the analysis of smart applications to enable IoTDots and the use of IoTDots Log Database (ITLD) for forensic purposes.

\subsection{Overview}
\begin{figure}[!t]
    \centering{\includegraphics[width=0.4\textwidth]{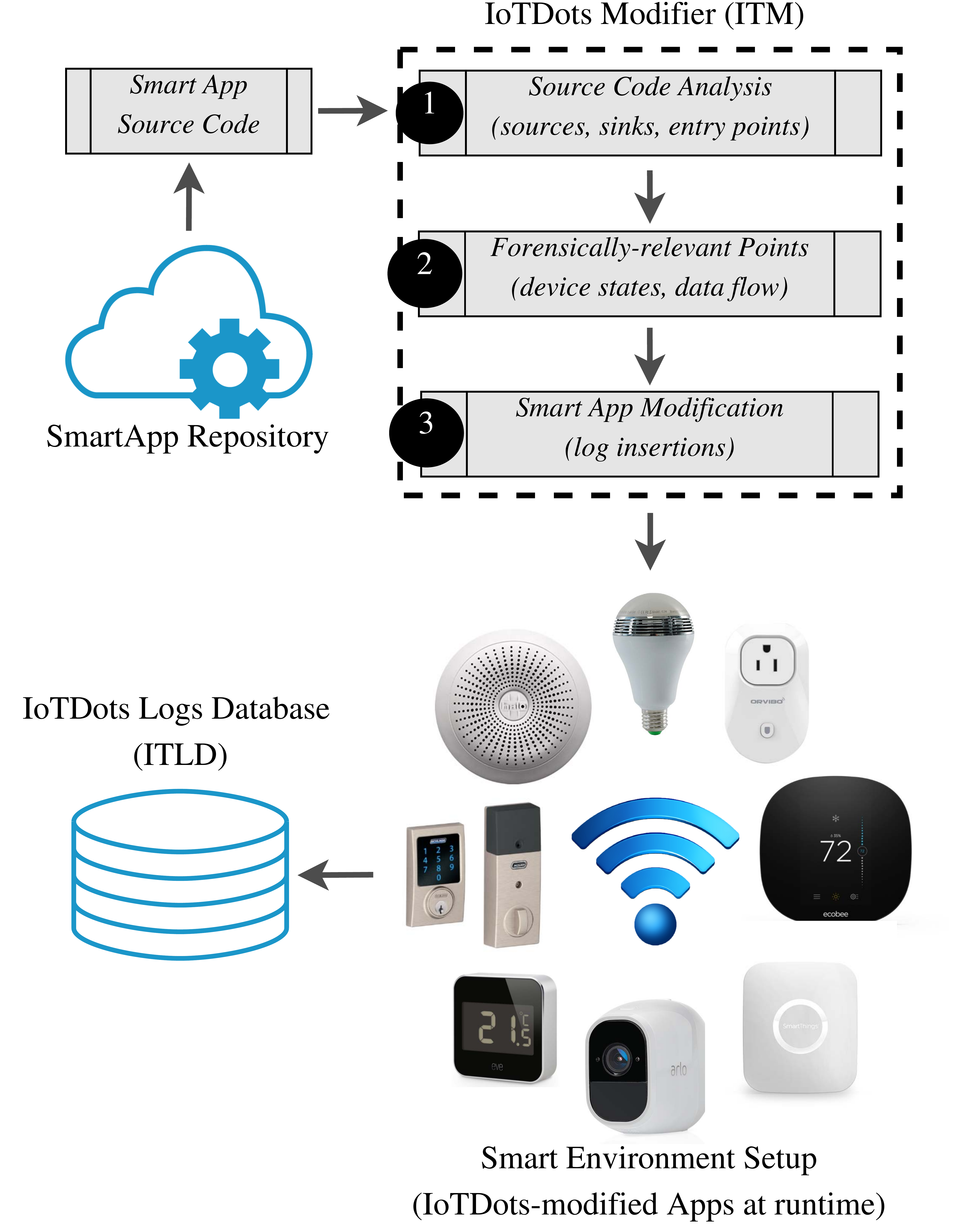}}
    \caption{The architecture of the ITM. Smart apps are analyzed to detect and log forensic-relevant data points at compile time. Then, the logs are sent to the ITLD at runtime.}
    \label{fig:Modifier}
\end{figure}

Our general architecture divides IoTDots into two parts: First, the IoTDots-Modifier (ITM) automatically analyzes and modifies the smart apps at compile time to allow forensically-relevant data to be sent to the ITLD in the form of forensic logs; second, logs from the IoTDots-modified smart apps stored in ITLD are processed by the IoTDots-Analyzer (ITA) to extract valuable forensic information. With this architecture, we envision the utilization of IoTDots in a corporate smart environment where only the use of IoTDots-modified smart apps is authorized. Figure \ref{fig:Modifier} shows the proposed architecture for the ITM. For the purpose of this work, we target open source IoT programming platforms since our implementation requires source code modification. Such modification requires that the ITM performs source code processing via static analysis on the smart apps' source codes directly extracted from the smart app repositories \cite{Official}. Hence, the ITM process includes (1) the analysis of the source code of the smart apps \cite{saint, SaINTWeb}, (2) forensically-relevant information detection, and (3) smart apps modification. Then, upon utilization, the IoTDots-modified apps send forensically-relevant logs containing states and actions occurring in the smart environment to the ITLD at runtime. Finally, the logs are organized and kept in the ITLD.

\begin{figure}[!h]
	\centering{\includegraphics[width=0.35\textwidth]{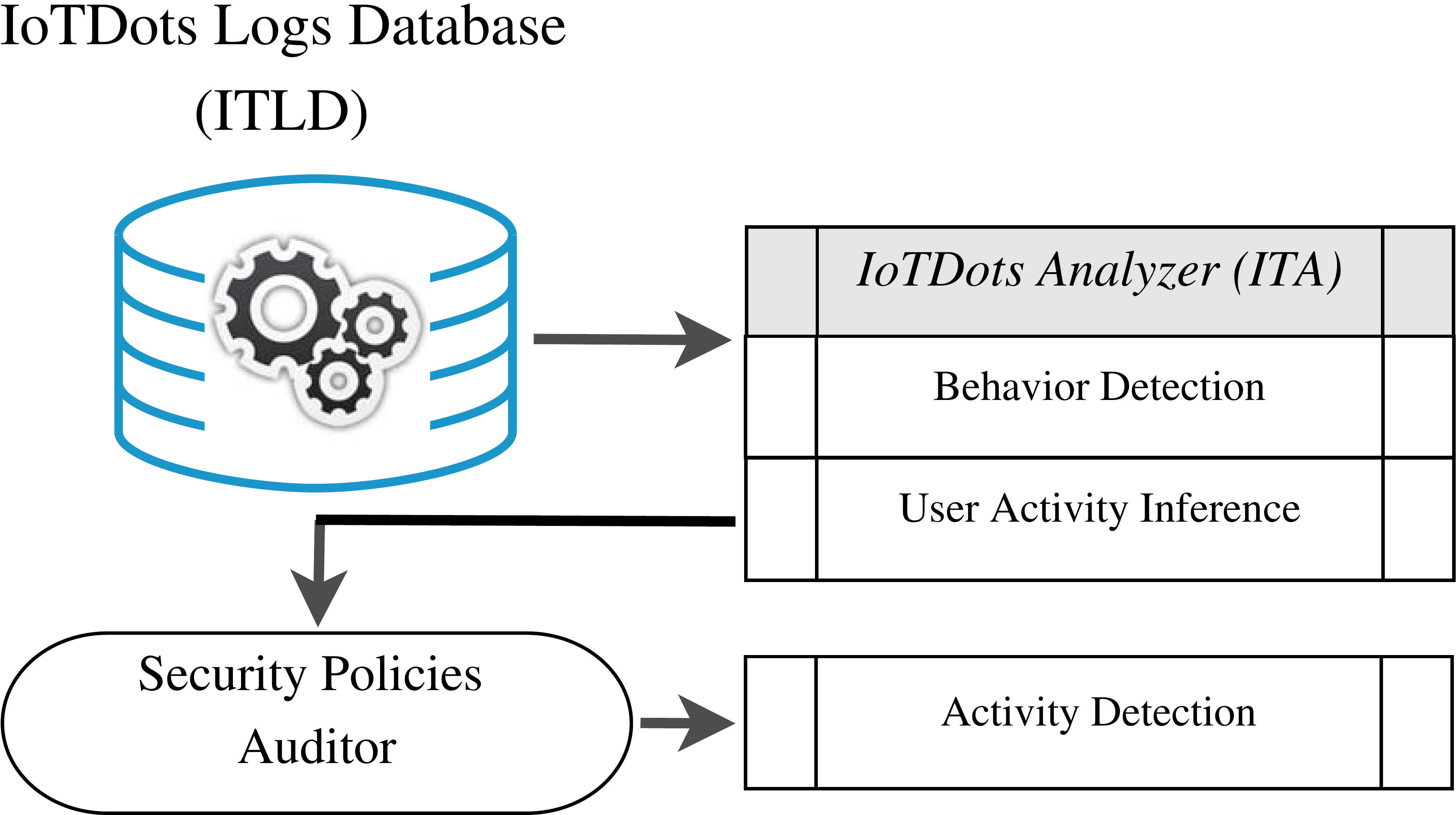}}
	\caption{The architecture for ITA. In the case of a forensic investigation, IoTDots logs are analyzed for potential forensically-valuable activities from users, apps, and devices in a smart environment.}
	\vspace{-0.2in}
	\label{fig:Analyzer}
\end{figure}

The second part of the architecture involves ITA, which is shown in Figure \ref{fig:Analyzer}. This stage performs data processing and analytics on IoTDots. The purpose of this analysis is to extract forensically-relevant information from the smart app logs. This information may potentially allow learning the state of the smart environment setup at any time. Also, this analysis provides insights into the users' activities occurring inside the smart environment. IoTDots is a framework capable of performing data processing and applying machine learning techniques to successfully combine logs from different smart devices, obtain a full description of the smart environment setup state, and learn all the possible activities that the users were performing inside the smart environment exactly at the time of interest of the forensic analysis. Finally, these activities are correlated with the security policies defined for the smart environment to detect user activities and potential malicious behaviors from users and smart apps. In the next sub-sections, we detail the important aspects of these operations.

\subsection{IoTDots Modifier (ITM)}
In this sub-section, we detail the ITM. As mentioned before, this part of the IoTDots framework analyzes the source code from the original smart applications to detect forensically-relevant information and automatically inserts specific logs for tracking purposes. Then, at runtime, the logs are stored in the ITLD where they are later analyzed for forensic purposes. In this work, we are targeting open source IoT programming platforms (e.g., Samsung SmartThings, OpenHAB, and Apple's HomeKit) which make the source code of smart apps available online \cite{Official, Community, openHABMarket, appleMarket}. However, before completing these steps, we first need to answer some key questions: 
What does the smart app life-cycle look like? How to define forensically-relevant information inside the smart app's source code? Where and how smart apps need to be modified to implement IoTDots? The following dissection of smart apps provides the answers to these questions. 

\subsubsection{Smart Apps Structure}
In general, smart app programming platforms \cite{smartThings-review, OpenHabGuideline, AppleHomekitReview} define the means to access and handle the apps' data and to transmit them out of the applications. Through these processes, smart applications can utilize device sensor readings, user-defined inputs, and events to execute the application logic either in the hub (i.e., hub-based setup) or in the cloud (i.e., cloud-based setup). Therefore, through the analysis of the smart applications' structure, one can identify and label smart app information \textit{points} that could potentially contain relevant data for forensic purposes. In the following, we describe five different smart app resources that are labeled by IoTDots as they potentially contain relevant information for forensic analysis. 

\begin{itemize}
\item\textit{Events}: These are used to respond to changes in the physical environment. At install time and depending on the app context, a smart application subscribes to a set of device events. Then, event handler methods are called every time such events occur. For forensic analysis, events are important since they define physical changes in the smart environment setup. In our analysis, we include device events as part of IoTDots' forensically-relevant information.
\item\textit{Actions}: After an event occurs, a smart app calls an action to control the affected device. Actions give ideas on how a smart environment setup respond to device changes. Often, these actions contain valuable timing information that can define changes in forensic timelines. 
\item\textit{User-defined Inputs}: Smart applications often require inputs from users either to manage the application logic or to control devices. These inputs are defined at install time and determine specific thresholds to trigger actions. Also, user inputs may define contact information to allow notification from smart apps to be sent to specific recipients. Changes in user-defined inputs are critical for forensic purposes since unauthorized modifications will directly impact the execution of the apps' events and actions and its notification process.
\item\textit{Device Information}: Applications also grant access to devices at install time to implement the application logic. Such devices complete the list of authorized smart entities in smart environment setups. Logging this information allows tracing authorized or unauthorized changes in the environment. 
\item\textit{Time and Location}: Successful forensic analysis of a smart environment  requires not only the timing information of events/actions, but also the physical location from where these events/actions were generated. In that way, such events can be directly attached to specific geo-location information of smart devices. 
\end{itemize}

The first step toward analyzing the smart app source code is to focus on the application's structure. In general, a smart app's structure follows the sensor-computation-actuator paradigm, which means, smart solutions are in general designed the same way regardless of their specific application and complexity. This common architecture highly simplifies the step of modeling the application structure from its source code. The benefits from modeling the smart app include the extraction of smart apps' entry points, events, and control flow of data. Also, it allows to identify entry/exit points of data (i.e., sources of data and sink functions) that are used to first define the origin of forensic-relevant information, and second how this information is sent out from the smart apps and to where.  

 \begin{figure*}[t]
    \centering{\includegraphics[width=0.9\textwidth]{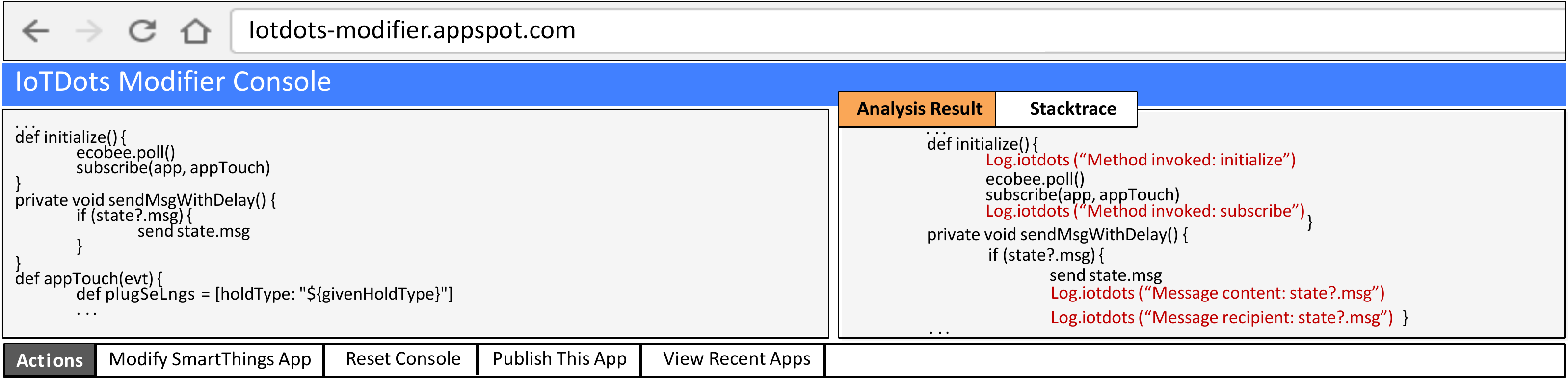}}
    \caption{We made IoTDots available online at https://iotdots-modifier.appspot.com/.}
    \vspace{-0.25in}
    \label{fig:webtool}
\end{figure*}

\subsubsection{Smart Apps Data Access Points}
Smart app programming platforms clearly define APIs \cite{SmartThingsAPI, OpenHABPrivacy} to transmit data outside the applications. These APIs can be categorized into two main groups:

\begin{itemize}
\item\textit{Internet}: Smart apps logic normally run as cloud services. This imposes a noticeable difference if compared to other domains (e.g., mobile) where local solutions are preferred. Therefore, smart apps are designed to directly act as web services or make calls to external services defined by the developers. These calls can make requests to smart apps via endpoints to obtain information such as device states, events, or even propose specific actions to control the smart solution. With IoTDots, we trace these requests through a custom logging mechanism.
\item\textit{Messages}: Smart applications can also define custom messages to send notifications to users. These notifications can be of the three different types: push notification in the mobile app, email, or short message services (SMS). This allows the developer to create a dedicated notification system to alert the user when specific events occur. Our framework also traces and stores these notifications for forensic purposes.  
\end{itemize}

\begin{lstlisting}[float=h!, caption= A sample code for a SmartThings App ,label=original]
/* A section of a code block of an original smart app */ 

section("Via a push notification and/or an SMS message") {
    input("recipients", "contact", title: "Send notifications to") {
        input "phone", "phone", title: "Enter a phone number to get SMS", required: false
    }
}
\end{lstlisting}

\begin{lstlisting}[float=h!, caption= A sample code for an IoTDots-modified SmartThings App ,label=modified, escapechar=!]
/* A section of a code block of an IoTDots-modified smart app */ 

section("Via a push notification and/or an SMS message") {
    input("recipients", "contact", title: "Send notifications to") {
        input "phone", "phone", title: "Enter a phone number to get SMS", required: false
    }
    !\colorbox{light-gray}{log.iotdots (\color{red}"New recipient defined: \${phone}"\color{black}) \color{blue}//IoTDots log}!
}
\end{lstlisting}

\subsubsection{Source Code and IoTDots Modifications of Smart Apps}
The purpose of the source code analysis of smart apps in IoTDots is to automatically detect the previously defined points of interest inside the source code and insert tracking logs for forensic purposes. The source code analysis starts by constructing an Inter-procedural Control Flow Graph (ICFG) of the apps and by extracting the nodes that define events, actions, user inputs, etc. For the specific purpose of this work, we target \textit{Samsung SmartThings} applications which are written in Groovy. Groovy is a Java-based programming language that defines \textit{visitors} for app methods and variables through the \texttt{ASTTransformationCollectorCodeVisitor} at compile time \cite{GroovyVisitor}. These visitors are then used to construct the Abstract Syntax Tree (AST) of the smart apps. The ITM contains a \textit{customizer} that (1) visits each node of the ICFG looking for forensic-relevant points and (2) modify the apps by inserting forensic logs in such points. 

In Listing \ref{original} and Listing \ref{modified}, we show how a smart application is modified by IoTDots for forensic purposes. After performing the source code analysis of the original app (see Listing \ref{original}), IoTDots flags a user-defined input \texttt{recipients}. This input defines the phone number to which all the notification from the smart apps will be sent to. As explained earlier in this section, these types of inputs are very critical for forensic purposes since careless or malicious users can modify this info anytime without being noticed. This can negatively impact the purpose of tracking the events and actions executed by apps being utilized in a smart environment. Going back to the previous example, after labeling the user input, IoTDots modifies the app by inserting a forensic log (see Listing \ref{modified}). Here, \texttt{log.iotdots} function defines an \texttt{http} request that pushes the log information to the remote ITLD. Then, the information is perpetually kept in the ITLD and can be accessed/processed anytime a forensic investigation request is received. 

Algorithm \ref{algo:iotdotsmodifier} details the processes performed by IoTDots to analyze and modify a smart app. During \textit{Detection}, the application source code is loaded into IoTDots modifier (Line \ref{line:ini1}). Recall that in this work, we are targeting open source smart app application platforms so availability of the source code is assumed. Then, the ICFG is calculated in Line \ref{line:icfg}. Once the ICFG is obtained, all the nodes are explored and all the forensic-relevant points are flagged in Line \ref{line:points}. This step concludes the Detection process in Algorithm \ref{algo:iotdotsmodifier}. Finally, the \textit{Modification} process customizes the smart apps by inserting the IoTDots logs in Line \ref{line:mod}.

\begin{algorithm}[h]
\footnotesize
\caption{Steps in the IoTDots-Modifier (ITM).}
 \label{algo:iotdotsmodifier}
 \begin{algorithmic}[1]

\STATE $appSC \gets$ app source code\\                            \label{line:ini1}

Detection:
\STATE $ICFG \gets$ createICFG($appSC$)                           \label{line:icfg}

\IF{Exists $ICFG$}                                                \label{line:ificfg}
    \FOR {nodes in $ICFG$}
        \STATE $forensicPT \gets$ forensic-relevant points        \label{line:points}
    \ENDFOR
\ENDIF \\
 Modification:
 \IF{$forensicPT$}        
     \FOR {points in $forensicPT$}
        \STATE Insert $IoTDots$ Logs                              \label{line:mod}
    \ENDFOR
\ENDIF
 \end{algorithmic}
 \end{algorithm}

Finally, we made ITM available online at:

\texttt{https://iotdots-modifier.appspot.com/}

The current version of the online tool automatically analyzes and modifies smart apps from Samsung SmartThings platform. We started implementing IoTDots for SmartThings apps mainly because (1) this platform defines the highest amount of different devices in the market, (2) it is open source; so the applications' source code can be found online, and (3) extensive API documentation is available online \cite{smartThings-documentation}. Figure \ref{fig:webtool} depicts details of the online version of ITM. On the left side, the user simply types or paste the original source code that needs to be modified to enable IoTDots. Then, on the right panel, the tool returns the modified app logging all the forensically-relevant source code information points.
 
\subsection{IoTDots Analyzer (ITA)}
ITLD stores logs obtained from smart apps at runtime. This allows the information from events and actions in a smart environment to be utilized later for forensic purposes. For successfully extracting information from the logs, we define ITA that performs the following actions on the ITLD data.
\begin{itemize}
    \item \textit{Labeling}: This step is used to label the data in the ITLD. Once classified, the data can be used to feed machine learning models to complete the analysis. The data processing step labels the ITLD data based on timing information and other forensic-related features. These features may include the type of devices generating the logs, location of these devices, etc.
    \item \textit{Detection}: One can think that the labeling of the ITLD data is good enough for forensic purposes. However, much more information can be extracted from the IoIDots logs. In this work, we introduce a framework not only capable of labeling the logs based on forensic criteria (i.e. timestamp, location, device), but also analyzing the data to infer user activity and detect the behavior of users and smart apps. For this purpose, ITA applies machine learning techniques to the ITLD-labeled data to classify and extract red-flagged forensic logs. Red-flagged logs can be used to define events and actions related to authorized user activity that change the smart environment topology (e.g., unauthorized device replacement or tampering, changes on user-defined inputs, unauthorized changes on setup topology, misplace or disable of any part of the smart setup, etc.). Also, red-flagged logs can be used by ITA to detect the abnormal or unauthorized behavior of users (e.g., users entering to access-limited zones of the building after work hours) or apps (e.g., malicious or tampered apps that force the smart setup to behave differently than expected). 
    \item \textit{Infer User Activity}: For the purpose of this work, \textit{user activity} is defined as any intentional or unintentional action performed by authorized users that cause changes in the smart setup and also violates the security policies established for the smart environment. This type of classification can hold users accountable for tampering or damaging the smart setup to avoid forensic logs.
    \item \textit{Detect Behavior}: For this work, \textit{behavior} of users or apps is defined as any user or app action that intentionally changes the purpose of the apps and also violates the security policies established for the smart environment. This type of classification hold users accountable of a possible misconduct or malicious activity or, on the contrary, will hold malicious or tampered smart applications responsible for making the smart setup to behave differently and potentially jeopardize the security of the smart environment. 
\end{itemize}
To further detail the operations of ITA, we introduce Algorithm \ref{algo:iotdotsanalyzer}. In Line \ref{line:vars}, $iotLogs$ variable is initialized with the content of ITLD. Additionally, the initialization step includes updating the variable $policy$ with the security policies that were valid for the smart environment at the time that the logs were acquired. Then, in Line \ref{line:labeling} the IoTDots logs are organized and labeled based on the timing information, the type of devices that generated the logs, and the location information of the devices. With this information, a machine learning model is applied on the data to (1) detect user actions in the smart environment (Line \ref{line:user}) and (2) detect behaviour of users and smart apps, all depending on the security policies established at the time the logs were acquired (Line \ref{line:policies}). Finally, if a forensic violation is detected, the flag is set to TRUE.

\begin{algorithm}[h]
\footnotesize
\caption{Steps in the IoTDots-Analyzer (ITA).}
 \label{algo:iotdotsanalyzer}
 \begin{algorithmic}[1]

\STATE $iotdLogs \gets$ ITLD\\                                    \label{line:vars}
\STATE $policy \gets$ smart environment security policies
\STATE $anomaly \gets$ FALSE

Labeling:
\FOR{each Log in $iotdLogs$}
   \STATE label Log by $Time$, $Device$, $Location$               \label{line:labeling}
   \STATE $MLdata \gets$ labels
\ENDFOR

Detection:
\FOR{data in $MLdata$}
    \STATE Evaluates user activity $userAct$                           \label{line:user}
    \STATE $ML \gets$ MLanalyzer($userAct$, $policy$)              \label{line:policies}
    \IF{Anomaly detected in $ML$}
        \STATE $anomaly \gets$ TRUE                                \label{line:anomaly}
    \ENDIF
\ENDFOR
 \end{algorithmic}
 \end{algorithm}

\section{IoTDots Forensic Evidence Detection} \label{sec:analysis}

In this section, we describe the analysis techniques that were used to identify user activities and behaviors in a smart environment from the collected forensic data. Specifically, we implemented a Markov Chain-based detection technique in IoTDots.

\subsection{IoTDots Data Characterization}
As noted earlier, IoTDots collects data from a smart environment in an ITLD that includes timing information, sensor information, device state, location, etc. For a specific time slot $t$, the data collected by IoTDots can be represented by:
\begin{equation}
Data\ array, E_t = \{S, D, M, L\},
\end{equation}
where $S$ represents the set of sensors' features, $D$ is the set of device features, $M$ is the features extracted from the controlling devices (smartphone/ tablet), and $L$ is the set of location features of controlling devices extracted from the log files. We describe the characteristics of these features below.

\begin{itemize}
\item\textit{Timing features (T):} A smart environment consists of several sensors and devices. These sensors and devices change their states based on different user activities and commands associated with the smart environment. In this context, some devices perform instantaneous tasks (e.g., switching lights with motion) while some devices perform a task over a period of time (e.g., changing temperature over time). 
IoTDots considers this timing information as a feature to infer the overall state of the smart environment at a specific time to detect the user activity and behavior.
\item\textit{Sensor features (S):} Sensors in a smart environment work as a trigger to different devices. A smart environment can have several different sensors (e.g., motion sensor, temperature sensor, presence sensor, etc.) connected to multiple devices. These sensors sense the changes in devices' peripheral and help the devices to take autonomous decisions such as switching lights on motion, triggering fire alarm after smoke, etc. Depending on the nature of the sensors, sensor data can be both logical (active/inactive) or numerical values. For IoTDots, we collect both numerical values and logical states of the sensors and create the state of the smart environment at a specific time. We represent the change in both logical state and a numerical value of a sensor as a binary input (1 if active/change and 0 otherwise) to create a forensic data matrix to train the detection algorithm.
\item\textit{Device features (D):} A smart environment supports different devices that are connected to a smart hub and different sensors and also with each other. These devices can perform multiple tasks as standalone devices (e.g., smart thermostat) or as connected devices (e.g., automatic door monitoring with smart lock and smart camera). For different user activities and input commands, these devices change their states (active/inactive) in an autonomous way. IoTDots collects these device state data (active/inactive state) from the stored log of the devices. 
\item\textit{Controller device features (M):} In a smart environment, users can use smartphones/tablets to control any device from the associated smart apps. IoTDots collects any control command generated from controller devices to understand user activities and on-going tasks in the smart environment.  
\item\textit{Location features (L):} The devices connected in a smart environment can be controlled from both inside and outside of the location. To understand the on-going activities in a smart environment, it is necessary to observe the exact location of a given command to the devices. IotDots considers the location of both the controller and the smart devices as a feature to understand any activities occurring in the smart environment. 
\end{itemize}

\subsection{Analytical Model Used in IoTDots}
ITA collects the information above and creates a state array to represent the state of the smart environment at specific time $t$. Each element of the state array represents the value of the features collected by IoTDots mentioned above. For a specific time $t$, we consider the combination of all the features collected by IoTDots as binary numbers (1 for active status and 0 for inactive status). Thus, the state of the smart environment can be represented as a n-bit binary number, where $n$ is the number of features extracted from the logs. Then, the total number of possible states of a smart environment with n number of features (sensors' states, devices' states, controller devices, and locations) can be $m=2^n$. IoTDots collects the state information of a smart environment over time to train a Markov Chain detection model to detect forensically-valuable behavior from users, smart apps, and devices in the system. The Markov Chain model has two basic assumptions which are as follows: (1) the occurrence probability of a specific state depends only on the previous one and (2) the transition between two consecutive states is independent of time and doesn't depend on any condition. Based on these assumptions, the Markov Chain model can be illustrated by the following equation~\cite{6thSense}.

\begin{equation}
\begin{split}
\begin{aligned}
P(X_{t+1} = x| X_1 = x_1, X_2 = x_2 ..., X_t = x_t) = 
\\P(X_{t+1} = x| X_t = x_t) , \\ 
when,\ P(X_1 = x_1, X_2 = x_2 ..., X_t = x_t) > 0,
\end{aligned}
\end{split}
\end{equation}
where $X_{t}$ and $X_{t+1}$ denotes state of the smart environment at time $t$ and $t+1$, respectively. Let assume the smart environment has the state $i$ at time $t$ and $j$ at time $t+1$. If $P_{ij}$ illustrates the transition probability between state $i$ to $j$, the state transition matrix for $m$ number of states of a smart environment can be represented by the following matrix.
\begin{equation}\label{transition}
P=\begin{bmatrix}
P_{11} & P_{12} & P_{13} & \ldots & \ldots & P_{1m}\\
P_{21} & P_{22} & P_{23} & \ldots & \ldots & P_{2m}\\
\ldots & \ldots & \ldots & \ldots & \ldots & \ldots \\
\ldots & \ldots & \ldots & \ldots & \ldots & \ldots \\
P_{m1} & P_{m2} & P_{m3} & \ldots & \ldots & P_{mm} \\
\end{bmatrix}
\end{equation}
To calculate each element of the transition matrix, let assume the smart environment has \textit{${X_0, X_1, \ldots, X_T}$} states at a given time \textit{${t= 0, 1, \ldots, T}$}. Then, each element of the transition matrix can be represented by the following equation~\cite{6thSense}:
\begin{equation}
{P_{ij}} = \frac{N_{ij}}{N_i}, 
\end{equation}
where $N_{ij}$ is the total number of transition from $X_t$ to $X_{t+1}$ over a certain period. From the state transition matrix, Markov Chain model can calculate the probability of occurring a state or sequence of states. To predict the probability of occurring a state, let assume the initial probability distribution of the Markov Chain model as follows:
\begin{equation}
Q = \begin{bmatrix}
q_1 & q_2 & q_3 & \ldots & \ldots & q_m,
\end{bmatrix}
\end{equation}
where, \textit{${q_m}$} is the probability that the model is in state \textit{m} at time 0. The probability of observing a sequence of states can be calculated by the following equation:
\begin{equation}
P(X_1, X_2, \ldots, X_T) = {q_{x1}} \prod_{2}^{T} {P_{{X_{t-1}}X{t}}}.
\end{equation}

\subsection{IoTDots Data Binarization}
Due to the nature of the IoTDots data, the acquired logs are not always Markov Chain-ready. In some cases, one needs to implement a binarizer that converts numeric data (e.g., sensor readings) into binary values that can be directly interpreted by the Markov Chain model. We found that, for the types of devices and sensors utilized in our evaluation, only a few cases of sensor readings require binarization. In these specific cases, IoTDots utilizes user-defined inputs in the smart apps to define threshold values that automatically convert numeric readings into binary values. For instance, for a temperature sensor, IotDots logs the sensor value for every device state change. Then, during analysis, these values are compared against the temperature value that the user set at install time. For sensor readings over the threshold, IoTDots feeds a value 1 to the Markov Chain model for that specific variable. On the other hand, for sensor readings below the threshold, IoTDots feeds a value 0 to the Markov Chain model. \textit{The use of user-defined inputs to implement the binarizer provides information to the IoTDots framework that may help to determine if the devices were compromised or were behaving in an unexpected way during the forensic incident}. Table \ref{tab:binarizer} summarizes the use of the binarizer for different types of data in IoTDots. 

\begin{table}[t]
\centering
\footnotesize
\begin{tabular}{cccc}
\toprule
\textbf{\begin{tabular}[c]{@{}c@{}}IoTDots Logs\end{tabular}} & \textbf{\begin{tabular}[c]{@{}c@{}}Data Type\end{tabular}} & \textbf{\begin{tabular}[c]{@{}c@{}}Requires\\Binarizer\end{tabular}} & \textbf{\begin{tabular}[c]{@{}c@{}}Comments\end{tabular}} \\ \hline
\midrule
\rowcolor{light-gray}
Location\tablefootnote{Refers to the Location Modes "Office" or "Other".} & Binary & \wcircle  & --  \\ 
Device & Binary  & \wcircle  & --  \\
\rowcolor{light-gray}
Sensor states & Binary  & \wcircle  & --  \\ 
Sensor values & Numeric  & \bcircle  & user-define inputs  \\
\rowcolor{light-gray}
Controller    & Binary  & \wcircle  & --  \\ 
\bottomrule
\end{tabular}
\caption{IoTDots implements a binarizer to convert numeric logs into binary values.}
\vspace{-0.3in}
\label{tab:binarizer}
\end{table}

We extract the state information of the smart environment from the collected logs and train the Markov Chain model. IoTDots then determines the on-going activities on the smart environment based on the state transition matrix explained in Equation~\ref{transition}. \textit{IoTDots predicts the next state of the environment and then match the predicted value with the state inferred from the dataset to determine if the current state has any forensic value or not}.

\section{Evaluation and Discussion} \label{sec:evaluation}

In this section, we evaluate the efficacy of IoTDots to detect forensic incidents such as regular and anomalous user activities as well as malicious user and smart app behaviors in a given smart environment. For a better analysis, we consider a hypothetical situation where a number of users perform regular activities inside a smart office environment while some others behave anomalously (e.g., by accessing office locations during unauthorized hours). Also, we consider another group of users that try to disable some smart devices to change the configuration of the smart environment and some malicious apps that poison the data from a specific number of devices. As detailed in Section \ref{sec:architecture}, our framework needs to first detect the forensically-valuable user activities to then classify them in regular or anomalous activities based on the security policies in place. In parallel, IoTDots has to be able to detect any forensically valuable behavior. We built the Markov Chain-based detection method for this purpose and train our model with data collected by IoTDots from a realistic smart office environment. Also, we evaluated the general overhead introduced by IoTDots to smart devices and cloud-based servers. 

\subsection{Training Environment and Data Collection}
To test the efficacy of the IoTDots performance, we collected daily usage data from a smart office environment. First, for training purposes, we implemented an emulator environment where day-to-day user activities could be simulated in a timely order. Then, the same activities were executed in real-life settings to collect log data from the real smart apps. 

\subsubsection{Training Environment Setup}
IoTDots is evaluated using smart apps and devices from \textit{Samsung SmartThings} platform. This allows performing tests using an IoT programming platform that defines the highest number of smart devices and has one of the largest market share~\cite{samsung}. Also, \textit{SmartThings} allows user-specific apps to be installed in a system which enables a multiple apps environment. We built a real-life setting with different devices and popular smart home apps available in the \textit{SmartThings App Market}. A detailed list of the different device types used in the experiments is given in Table~\ref{tab:sensors}.

\begin{table}[h!]
\centering
\footnotesize
\begin{tabular}{cc}
\toprule
\textbf{\begin{tabular}[c]{@{}c@{}}Device \\ Type\end{tabular}} &  \textbf{Model}\\ \hline
\midrule
Smart Home Hub & Samsung SamrtThings Hub \\\hline. 
Smart Light & Philips Hue Light Bulb \\\hline
Smart Lock & \begin{tabular}[c]{@{}c@{}}Yale B1L Lock with \\Z-Wave Push Button\\Deadbolt\end{tabular}\\\hline
Fire Alarm & \begin{tabular}[c]{@{}c@{}}First Alert 2-in-1 \\Z-Wave Smoke Detector and \\Carbon Monoxide Alarm \end{tabular} \\\hline
Smart Monitoring System & \begin{tabular}[c]{@{}c@{}}Arlo by NETGEAR\\ Security System \end{tabular}\\\hline
Smart Thermostat & Ecobee 4 Smart Thermostat \\\hline
\begin{tabular}[c]{@{}c@{}}Motion Sensor\\Light Sensor\\Temperature Sensor\end{tabular} & Fibaro FGMS-001 Motion Sensor\\\hline
Door Sensor & Samsung Multipurpose Sensor\\
\bottomrule
\end{tabular}
\caption{List of smart devices used during the experiments.}
\label{tab:sensors} 
\vspace{-0.5cm}
\end{table}

Our set of forensically-valuable activities includes the following scenarios:
\begin{itemize}
    \item \textit{Time-dependent access}: We provide user access to all the devices during office hours to observe user activities in the smart environment. 
    \item \textit{Restricted access}: For specific places in the smart environment, we enforce restricted access policies.
    \item \textit{Multi-user environment}: We consider a multi-user environment where different users perform different activities inside the smart environment. In a multi-user scenario, we also consider time-dependent access and restricted access to emulate the real-life smart office environment. 
\end{itemize}

\begin{figure*}[t]
\centering
    \subfigure[]{\includegraphics[width=0.30\textwidth]{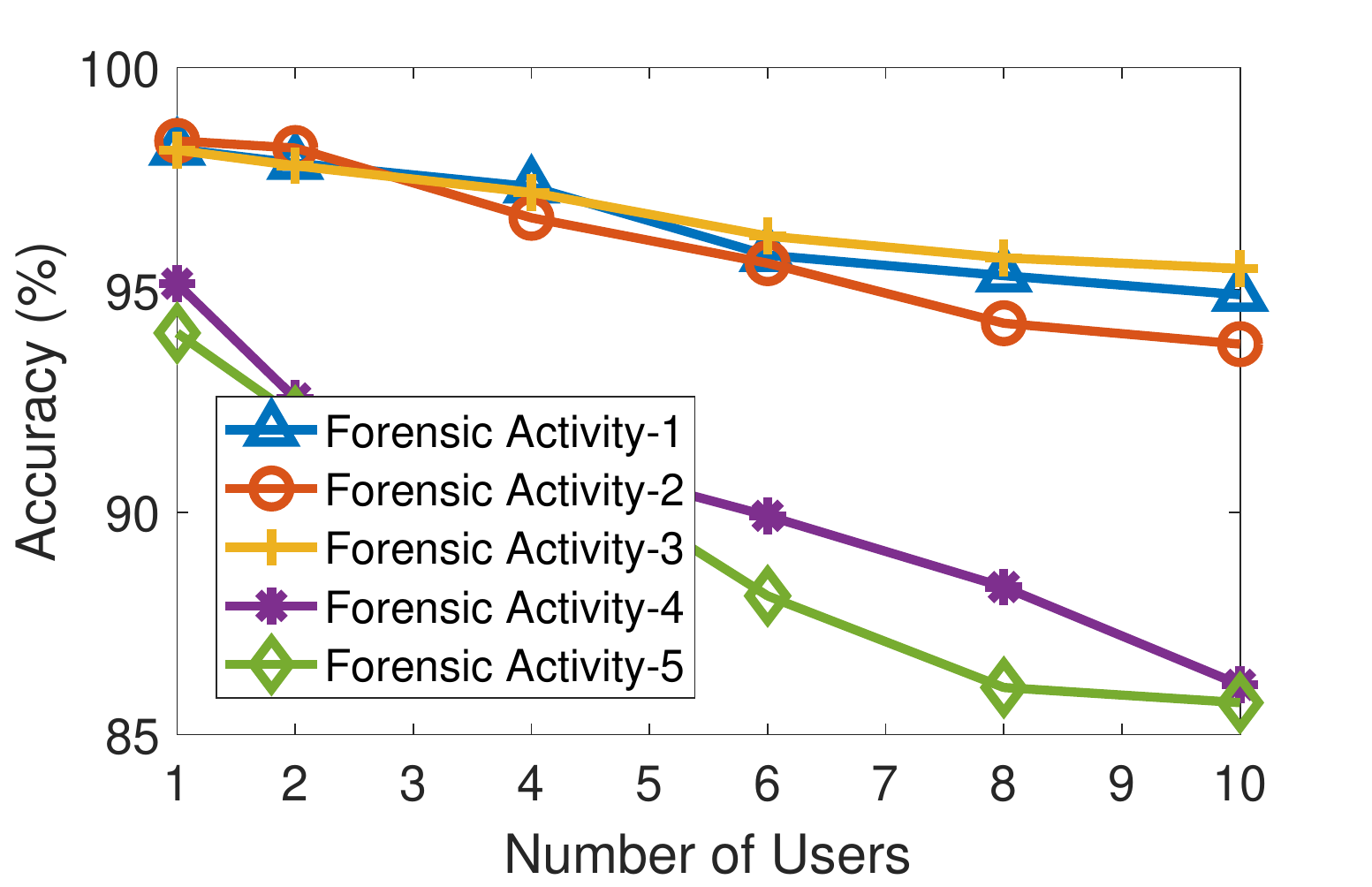}\label{mult-user}}
    \subfigure[]{\includegraphics[width=0.30\textwidth]{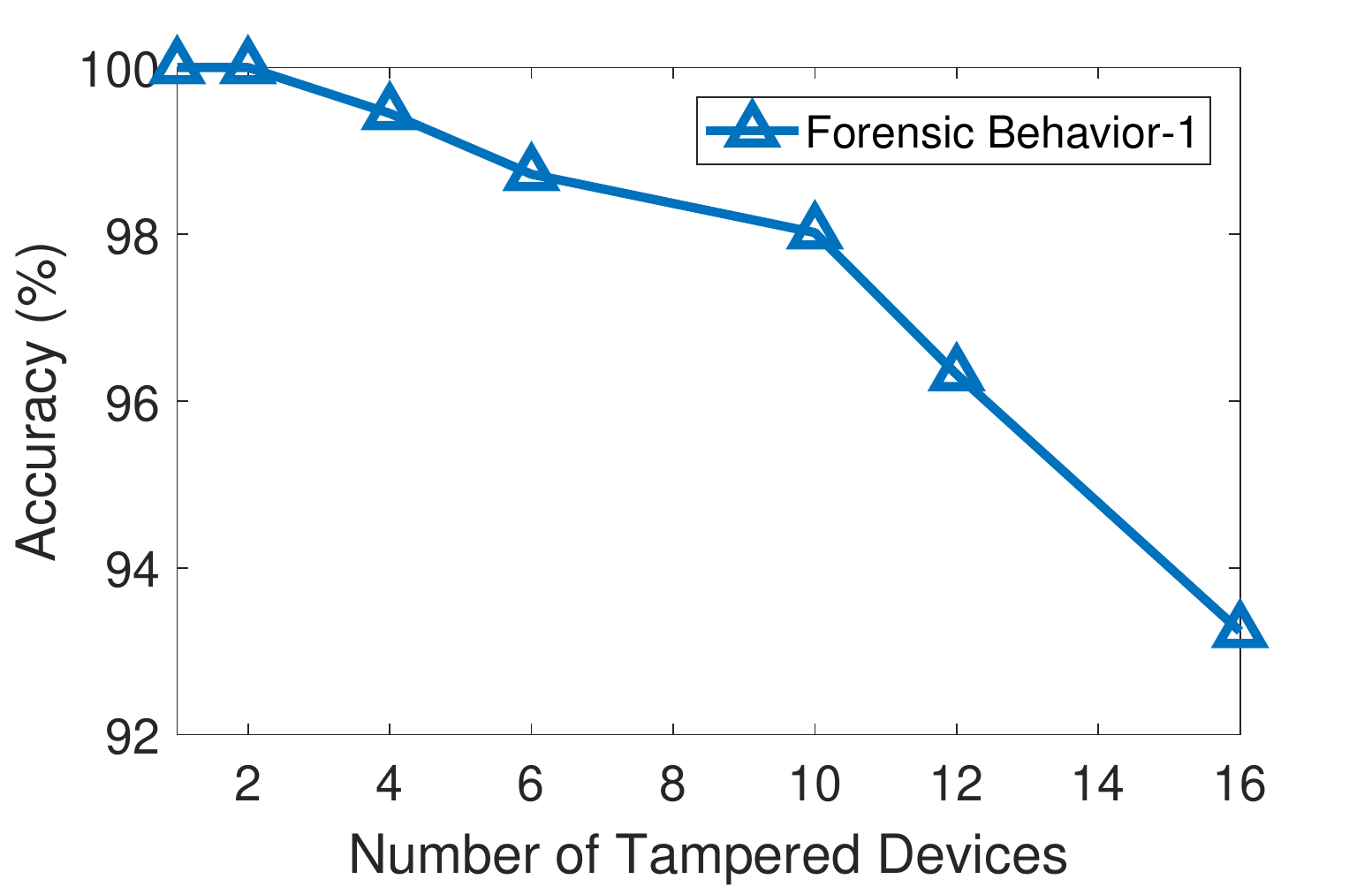}\label{tampered}}
    \subfigure[]{\includegraphics[width=0.30\textwidth]{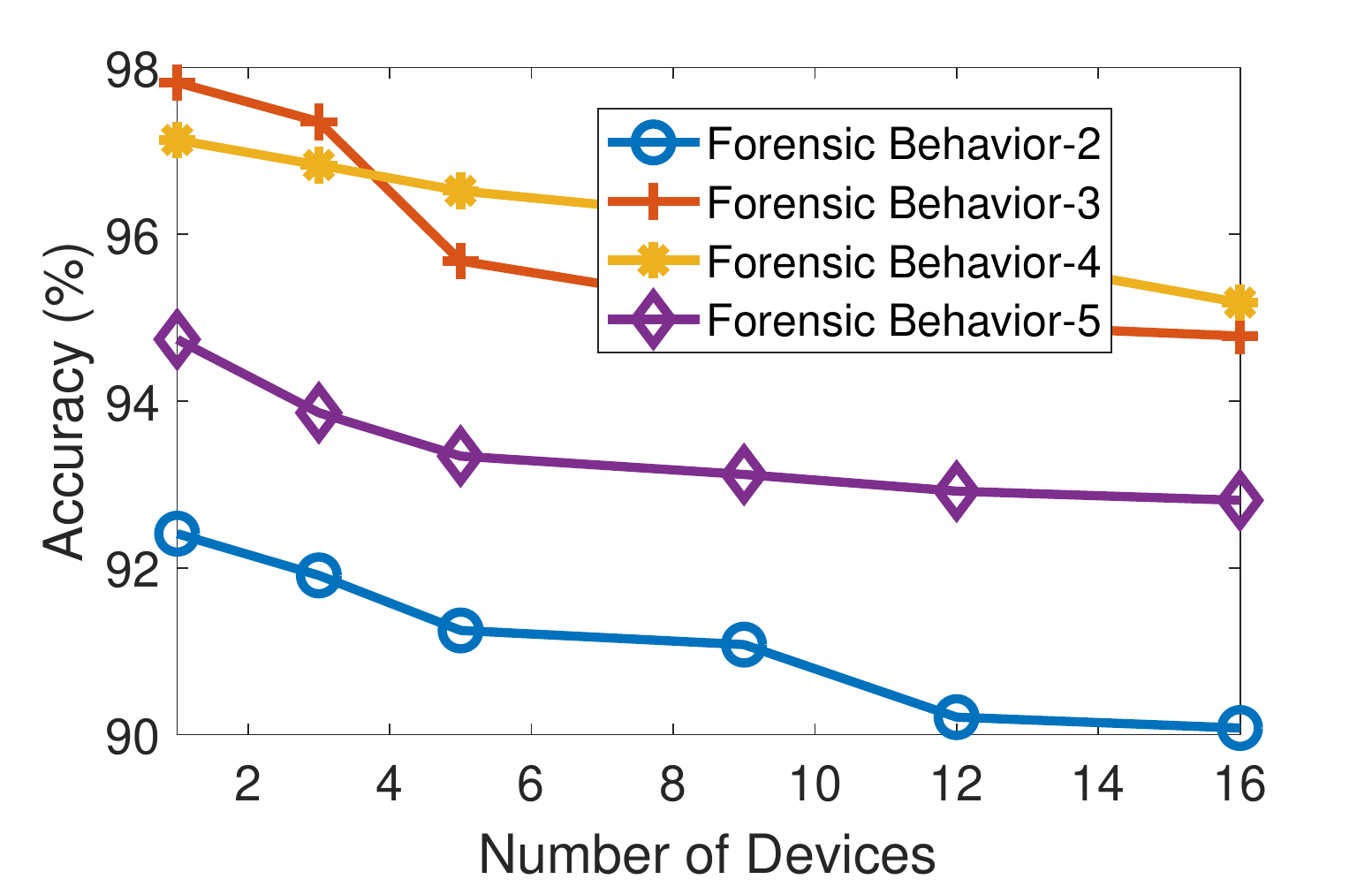}\label{threats}}
    \vspace{-0.05in}
    \caption{IoTDots evaluation results for different forensic incidents: (a) Accuracy of IoTDots on inferring activities in multi-user scenarios, (b) accuracy of IoTDots on detecting forensic behavior from users versus the number of tampered devices, and (c) accuracy of IoTDots on detecting forensic behavior from smart apps versus different number of devices. The different curves represent the performance of IoTDots for activities and behaviour as described in Section \ref{sec:threatmodel}.}
    \vspace{-0.1in}
\end{figure*}

\subsubsection{Data Collection}
To enable the data collection, we used ITM to automatically modify the SmartThing apps by performing source code analysis to detect forensically-relevant information inside the source code and insert specific logs. Then, while utilizing the apps, IoTDots data was sent to the ITLD. We collected data from the interaction of 10 different users and 22 smart devices (10 different types of devices and sensors, as shown in Table \ref{tab:sensors}) and sensors for seven days for a total of 84209 forensically-valuable data incidents. Our dataset included data collected from benign and malicious environments. For the first case, the users were allowed to perform anomalous activities in clear violation of the security policies defined for our experiments. These anomalous activities are described in Section ~\ref{sec:threatmodel}. In total, we collected benign forensic data from 30 different experiments with over 3000 instances for five different forensically-valuable anomalous behaviors. 

For collecting data from a malicious forensic environment, we considered two different cases with users and smart apps. For the case of users, we considered users trying to modify, tamper, remove, or destroy the devices with the intention of changing the original setup of the smart environment and prevent that forensic logs can be sent to the ITLD for future analysis. On the other hand, for the apps, we installed different popular \textit{SmartThings} apps along with multiple malicious apps \cite{iotbench}. We created four different forensic scenarios (see Section~\ref{sec:threatmodel}) and built their corresponding smart apps similar to \cite{saint, ContexIoT-repo} based on the threat model explained in Section~\ref{sec:threatmodel}. For Threat 2, we created two different apps for a smart lock that leaked the access code to an unauthorized person. This also represents device tampering by authorized users. For Threat 3, we built an app that injected forged data in a fire alarm and triggered the alarm maliciously. For Threat 4, we developed an app that could shut down the smart thermostat for a specific input temperature. Finally, for Threat 5, we created an app for a smart camera that could be triggered by a specific light pattern. In total, for these different scenarios, we collected 50 different forensic datasets to test the effectiveness of IoTDots. 

We also note that we used 75\% of the user data to train the Markov Chain model and then combined the rest 25\% of the data along with malicious data to test ITA. For performance metrics, we utilized standard parameter, including accuracy, F-score, True Positive Rate (TPR), False Positive Rate (FPR), True Negative Rate (TNR), and False Negative Rate (FNR).  

\subsection{Forensic Activity Detection from Users} 
The state of the interconnected devices inside the smart environment depends on the on-going user activities. For example, while a user moves from one place to another, several devices and sensors become active. The changes on device states can be an instantaneous event (a specific event at a specific time such as switching on a light) or a combination of sequential events over a period of time (motion from one place to another). 

\begin{table}[h!]
\centering
\footnotesize
\begin{tabular}{ccccccc}
\toprule
\textbf{\begin{tabular}[c]{@{}c@{}}User \\Activity\end{tabular}} & \textbf{\begin{tabular}[c]{@{}c@{}}TPR\end{tabular}} & \textbf{\begin{tabular}[c]{@{}c@{}}FNR\end{tabular}} & \textbf{\begin{tabular}[c]{@{}c@{}}TNR\end{tabular}}   & \textbf{\begin{tabular}[c]{@{}c@{}}FPR\end{tabular}} & \textbf{ACC} & \textbf{\begin{tabular}[c]{@{}c@{}}F-\\Score\end{tabular}}\\ \hline
\midrule
\rowcolor{light-gray}
Activity-1  & 0.9914  & 0.0086  & 0.9371  & 0.0629  & 0.9823  & 0.9635  \\ 
Activity-2  & 0.9926  & 0.0074  & 0.9334  & 0.0666  & 0.9874  & 0.9621  \\ 
\rowcolor{light-gray}
Activity-3  & 0.9886  & 0.0114  & 0.9619  & 0.0381  & 0.9830  & 0.9750  \\ 
Activity-4 & 0.9687   & 0.0313  & 0.8714  & 0.1286  & 0.9484  & 0.9175  \\
\rowcolor{light-gray}
Activity-5 & 0.9555   & 0.0445  & 0.8708  & 0.1292  & 0.9388  & 0.9112  \\
\bottomrule
\end{tabular}
\caption{Performance evaluation of IoTDots for inferring forensically-valuable user activities.}
\vspace{-0.2in}
\label{fig:activities}
\end{table}

Table~\ref{fig:activities} illustrates the detailed evaluation of IoTDots for user activity inference. One can observe that for time-independent actions (i.e., Activity 1-3), IoTDots can achieve both accuracy (i.e., ACC) and F-score over 98\% and 96\%, respectively. True Positive Rate (TPR) and True Negative Rate (TNR) are also high (over 98\% and 93\%, respectively). On the other hand, for time-dependent actions (i.e., Activity 4-5), IoTDots achieves the highest accuracy and F-score of 94.84\% and 91.75\%, respectively. Since the time-dependent actions are related to user motions, different users may have different patterns of motion which increases the False Positive rate (FPR) and False Negative rate (FNR). In summary, IoTDots achieves over 93\% of accuracy for detecting different user activities forensically. User action inference in the smart environment also depends on the number of users present in the environment. 

For the case of a multi-user smart environment, IoTDots requires a different analysis since users can perform different tasks at once which can directly impact the accuracy of the user action inference. In Figure~\ref{mult-user}, the accuracy of forensic action inference is shown with respect to the number of present users in the smart environment. As analyzed, it can be observed how the accuracy values decrease with the users. From Figure~\ref{mult-user}, one can observe that for time-independent activities (Activity-1, Activity-2, Activity-3), IoTDots achieves accuracy in the range of 98\% to 95\%. For time-dependent activities (Activity-4, Activity-5), the accuracy of IoTDots varies from 95\% to 86\% as the number of users increases.

\subsection{Detection of Forensic Behavior from Users}
Smart devices installed in a smart environment are vulnerable to device tampering which can lead to malicious events. In previous works that proposed the use of IoT data for forensic investigations, device integrity was always assumed and the results were vulnerable to insiders \cite{fear}. IoTDots can detect tampered devices based on the collected logs from the smart environments. For this, the Markov Chain model analyses the state of all the devices in the smart environment at any given time and detects malicious or unexpected states by comparing data from similar devices sharing the same context (i.e., device cooperation). During device cooperation, if one device is compromised or tampered, the information collected from other trusted devices inside the environment is used to detect the one reporting fake data. On the other hand, if the majority of the devices are compromised, the overall system cannot be trusted which will impact the evaluation. For example, consider a smart light that is controlled by a motion sensor. During normal operations, if the sensor detects any motion, the light will be on. However, in the event of a compromised state of the smart light, the light will not operate properly with the motion sensor. On the other hand, if the motion sensor is compromised too, it would be hard to define which device is hampering the normal operation of the devices. Figure~\ref{tampered} depicts the accuracy of IoTDots on detecting tampered or modified devices in a smart environment. For this, we installed 22 different devices (including smart sensors) in a smart office environment. One can notice from Figure~\ref{tampered} that IoTDots can achieve near 100\% of accuracy on cases with 2 tampered devices in the environment. In general, the accuracy of IoTDots decreases as the number of tampered devices increase in the system. Finally, Figure \ref{tampered} demonstrate how device cooperation is affected by the number of compromised devices. One can observe how the accuracy values drop when more than 10 devices (near 50\% of the total number of devices) are compromised.

\subsection{Detection of Forensic Behavior from Smart Apps}
IoT programming platforms offer customized apps to control the smart devices. In recent years, researchers have reported several malicious apps that can perform malicious activities in a smart environment~\cite{fernandes2016security, smartTV-hack}. In this section, we test the efficacy of IoTDots in detecting behaviors in a smart environment caused by apps installed on the devices. As noted earlier, we consider four different scenarios to evaluate app behavior in IotDots (Section \ref{sec:threatmodel}). To evaluate these scenarios, we installed malicious IoTDots-modified apps in a real-life smart environment (smart office). Table~\ref{fig:threats} shows the evaluation of IoTDots in detecting app's behavior in the smart environment. One can observe that IoTDots achieves high accuracy and F-score for all the aforementioned cases (over 95\% and 89\%, respectively). 
\begin{table}[h!]
\centering
\footnotesize
\begin{tabular}{ccccccc}
\toprule
\textbf{\begin{tabular}[c]{@{}c@{}}Behavioral \\Model\end{tabular}} & \textbf{\begin{tabular}[c]{@{}c@{}}TPR\end{tabular}} & \textbf{\begin{tabular}[c]{@{}c@{}}FNR\end{tabular}} & \textbf{\begin{tabular}[c]{@{}c@{}}TNR\end{tabular}}   & \textbf{\begin{tabular}[c]{@{}c@{}}FPR\end{tabular}} & \textbf{ACC} &
\textbf{\begin{tabular}[c]{@{}c@{}}F-\\Score\end{tabular}}\\ \hline
\midrule
\rowcolor{light-gray}
Behavior-2  & 0.9564  & 0.0436  & 0.8479  & 0.1521  & 0.9468  & 0.8989  \\ 
Behavior-3  & 0.9669  & 0.0331  & 0.9333  & 0.0667  & 0.9636  & 0.9498  \\
\rowcolor{light-gray}
Behavior-4  & 0.9725  & 0.0275  & 0.9289  & 0.0711  & 0.9670  & 0.9502  \\ 
Behavior-5  & 0.9647  & 0.0353  & 0.8953  & 0.1047  & 0.9572  & 0.9287  \\
\bottomrule
\end{tabular}
\caption{Performance evaluation of IoTDots in detecting forenscially valuable behavior from smart apps.}
\vspace{-0.2in}
\label{fig:threats}
\end{table}

In Figure~\ref{threats}, the accuracy of IoTDots is shown for different forensic cases with respect to the number of devices presented in the environment. IoTDots achieves the highest accuracy of over 97\% for Behavior-3 and the lowest accuracy of 92.3\% for Behavior-2 in the case of only one device installed in the system. With the increase in the number of devices, the accuracy decreases to 95\% and 90\% for Behavior-3 and Behavior-2, respectively. The accuracy of detecting Behavior-4 and Behavior-5 varies between 97\% to 95\% and 94.8\% to 93\% with the number of devices in the system, respectively. 

\begin{figure}[!h]
\centering{\includegraphics[width=0.35\textwidth]{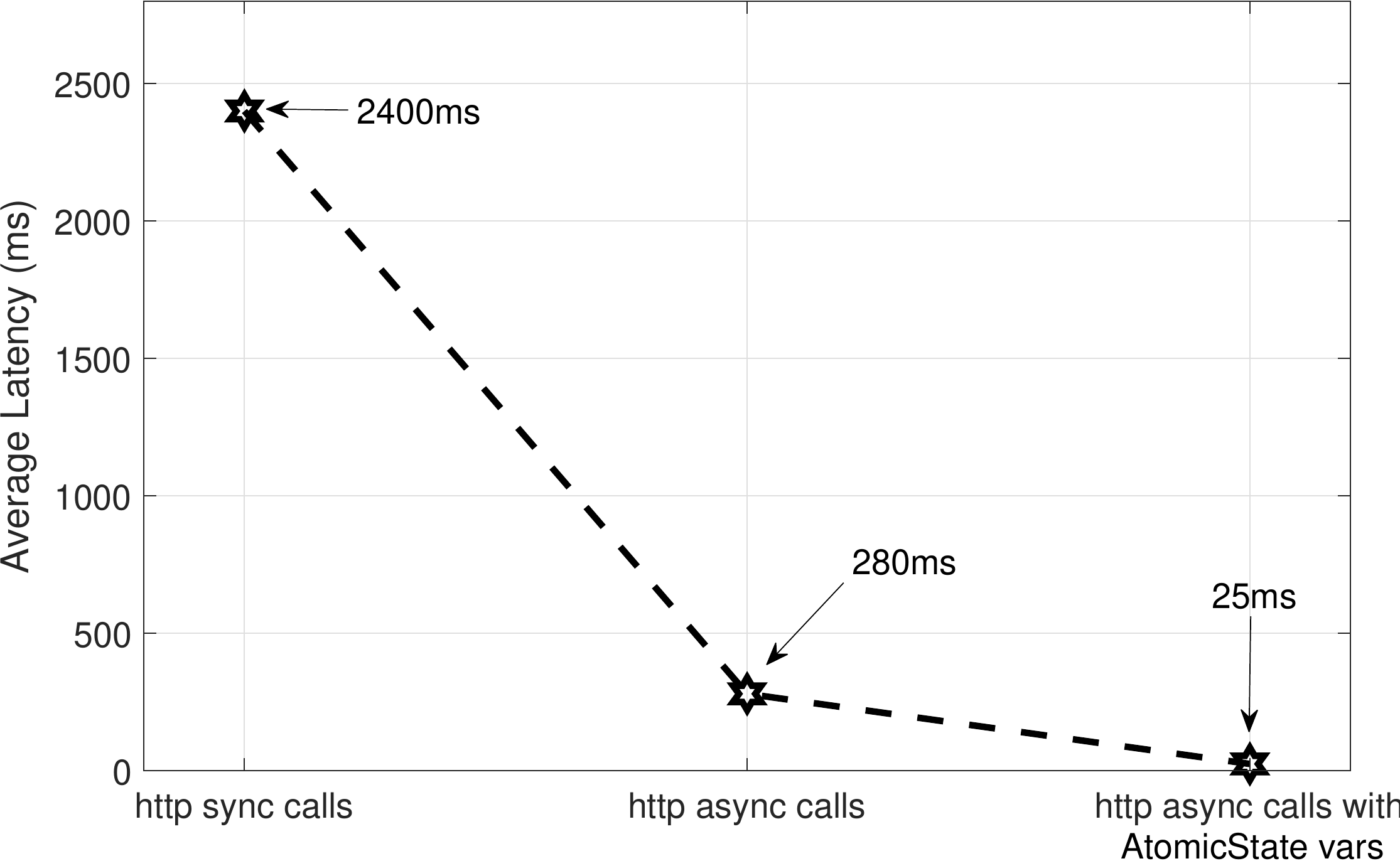}}
    \caption{Average latency imposed by IoTDots to smart apps' execution times. The maximum latency (2.4s) is obtained when synchronous HTTP requests are utilized to send the logs to ITLD. Then, the latency is significantly reduced to 280 ms after using the asynchronous HTTP requests. Finally, the minimum latency (25ms) is obtained after combining the asynchronous HTTP request with AtomicState queuing.}
\label{fig:latency}
\end{figure}

\subsection{System Overhead}
Since most of the smart apps are cloud-based, we do not expect that IoTDots impose any overhead on the smart devices. However, it is necessary to evaluate the impact of IoTDots on the computing resources of the IoT smart apps servers in the cloud. To assess this, we selected two metrics that are directly related with (1) the amount of physical memory occupied by the IoTDots-modified apps and (2) the latency of the smart apps in their operations and responses. 

\noindent \textbf{Physical Memory}: Since server space is costly, we need to evaluate how much more physical space the IoTDots-modified apps require from the cloud servers if compared with the original smart apps (previous to IoTDots modification). On average, ITM adds around 110 new lines of code to the original smart apps source code, which represent an increase of around only 5KB of physical memory space per app.

\noindent \textbf{Latency}: For the purpose of this work, we define \textit{latency} as the extra delay imposed by the IoTDots-modified apps source code to the original execution time of the smart apps. In general, the latency depends on (1) the number of times that the IoTDots logs are sent to the cloud and (2) the IOTDots log function execution time. In terms of latency, the worst case scenario is obtained when an individual synchronous \texttt{http} request is called for every single IoTDots log. In that case, around 2.4s latency is added every time a log is sent from the app to ITLD, which represents a big overhead to the app operation. To avoid this latency, IoTDots utilizes asynchronous HTTP request to send data to ITDL. Since we are evaluating apps from the Samsung Smartthings platform, we utilized the beta version of the \texttt{$asynchttp\_v1$} class \cite{smartThings-documentation} to asynchronously send the logs to the database. Further, we reduced the latency overhead even more by avoiding \texttt{http} requests every time a new log is generated. For this, we utilized the \texttt{AtomicState} variable in SmartThings to queue and map several logs together before communicating with the ITDL. Finally, with these modifications, the average latency overhead imposed by IoTDots is reduced to around 30 ms for every \texttt{http} request to the ITDL (see Figure \ref{fig:latency}).  

\subsection{Summary}
\textit{Overall, our evaluation results showed that IoTDots can infer and classify forensically valuable user activities by considering changes in sensor reading and device states. For inferring such activities, IoTDots pairs user activities with the security policies defined for the smart environment to detect the activities. This may help the forensic investigators to know what was happening before, during, and after the time of the analyzed incidents. Furthermore, forensic results can be extended by considering the sequence of the different device states in the smart environment over time, which may help to create a timeline of the incidents. Our results showed that forensically-valuable information can be easily detected with high accuracy using the previously mentioned features. This may help the investigator to further analyze if incidents were caused due to users' negligence or due to specific activities (either benign or malicious).}

\begin{table*}[t]
\centering
\small
\begin{tabular}{ccccccc}
\toprule
\textbf{\begin{tabular}[c]{@{}c@{}}Tool Name\end{tabular}} & \textbf{\begin{tabular}[c]{@{}c@{}}IFA\end{tabular}} & \textbf{\begin{tabular}[c]{@{}c@{}}Cross\\ App \\Analysis\end{tabular}} & \textbf{\begin{tabular}[c]{@{}c@{}}Consider \\Devices\end{tabular}}   & \textbf{\begin{tabular}[c]{@{}c@{}}No Platform \\Modification\end{tabular}} & \textbf{\begin{tabular}[c]{@{}c@{}}Consider \\Tampered\\Devices\end{tabular}} & \textbf{Goals \& Comments} \\ \hline
\midrule
\rowcolor{light-gray}
FlowFence  & \bcircle  & \bcircle  & \wcircle  & \wcircle  & \wcircle   & \begin{tabular}[c]{@{}c@{}}Protects data by enforcing data flow policies from users. \end{tabular}  \\ 
ContextIoT & \bcircle  & \wcircle  & \wcircle  & \bcircle  & \wcircle   & \begin{tabular}[c]{@{}c@{}}Detects malicious data flow by analyzing app \\context in a simulation environment.\end{tabular}  \\
\rowcolor{light-gray}
ProvThings & \bcircle  & \bcircle  & \bcircle  & \bcircle  & \wcircle   & \begin{tabular}[c]{@{}c@{}}Detects malicious data flow by analyzing app context in \\real environments. Considers app-device and device-device \\interactions.\end{tabular}  \\ 
SaINT      & \bcircle  & \wcircle  & \bcircle  & \bcircle  & \wcircle   & \begin{tabular}[c]{@{}c@{}}Detects data exflitrations from smart apps. \end{tabular}  \\
\rowcolor{light-gray}
IoTDots    & \bcircle  & \bcircle  & \bcircle  & \bcircle  & \bcircle   & \begin{tabular}[c]{@{}c@{}}Enables forensics analysis in smart environments.\\ Assumes device-device, app-device, and user-device \\interactions. Considers tampered devices.  \end{tabular}  \\ 
\bottomrule
\end{tabular}
\caption{Comparison between IoTDots and other IFA tools.}
\label{tab:IFA}
\vspace{-0.3cm}
\end{table*}

\section{Benefits, Challenges, Limitations, and Future Work} \label{sec:benefits}
There are several benefits associated with the use of IoTDots. Also, some challenges and limitations can be highlighted.
\begin{enumerate}
    \item \textit{Forensic Framework for Smart Environments}: Smart environments are equipped with a myriad of different smart devices and sensors. This represents a new domain of relevant data that can be used for forensic purposes. In this context, IoTDots emerges as the first practical solution that collects, stores, and processes smart environment data to instrument forensic analysis. 
    \item \textit{Automatic Log Insertion}: IoTDots-Modifier automatically analyzes smart apps source code to identify and log forensic-relevant data. Then, the logs are sent to a database where the data is kept for future analysis. The current version of IoTDots performs automatic analysis and modification for Samsung SmartThings apps only. The new version with support to other IoT programming platforms is currently under development. Despite most IoT apps follow similar architectural paradigms, differences in the programming languages and other platform-specific features make the expansion of IoTDots to other platforms a challenge. 
    \item \textit{Deep Data Analysis}: IoTDots-Analyzer incorporates data processing and machine learning techniques to label the IoTDots data and detect forensically-valuable anomalous activity and behavior from users, smart apps, and devices in a smart environment. The current version has shown excellent results on accuracy metrics for different cases of numbers of users and devices. However, analysis of new evaluation cases may be necessary. For instance, different machine learning algorithms may be considered to avoid conversion of the IoT data into Markov binary states.
    \item \textit{Low overhead}: IoTDots imposes minimal to no overhead to the devices deployed in the smart environment and very low overhead to the cloud servers supporting the smart apps. In general, IoTDots-modified apps contain around 110 lines more on average if compared with the original smart apps. Also, the latency imposed by sending the IoTDots logs to the cloud is very low. 
    \item \textit{Applicability}: The approach used in IoTDots can be easily generalized to any IoT platform. For the current version, IoTDots takes advantage of open source platforms to enable simplicity. Closed-source IoT systems like HomeSeer \cite{homeseer} and HomeOS \cite{homeos} may require modification on ITM to allow source code analysis of the apps and the implementation of an effective logging system. 
    \end{enumerate}

As future work, we aim to incorporate new smart app platforms to IoTDots. Since different platforms may use different programming languages and APIs, the ITM may require some architectural modifications to enable source code analysis. Also, future versions of our framework will incorporate new threats to the analyzer as new types of devices and user behaviors are considered in the smart environment. 

\section{Related Work} \label{sec:related}
Smart environments such as smart home systems have become popular in recent years. However, there exists no comprehensive platform for forensic analysis of data from a smart environment that considers the interaction between smart devices, users, and the security policies in place. In this section, we discuss existing forensic data analysis platforms for smart environments and their shortcomings.

\subsubsection{Forensic Data Collection from the Smart Environment}
Researchers have proposed different approaches to collect forensic data from smart environment and smart devices. Some of these works only focus on vendor-specific devices~\cite{chung2012digital}, while others present general methods to only collect forensic data without any data analysis option~\cite{oriwoh2013internet, watson2016digital}. Kebande et al. proposed a generic approach, \textit{DFIF-IoT}, to analyze digital forensic data in IoT settings~\cite{kebande2016generic}. Here, the authors presented a generalized method to capture forensic data from IoT devices including cloud, network, and device-level forensic data. However, \textit{DFIF-IoT} only discusses the theoretical approach to collect forensic data from IoT environment without any real-life implementation and evaluation, which decreases the practicality of this work. Zawoad et al. proposed FAIoT, a forensic-aware eco-system to collect forensic data from IoT platforms~\cite{zawoad2015faiot}. FAIoT presents a general platform to collect forensic data in a systematic way and organizes data for further analysis. One shortcoming of this work is that there is no implementation of data analysis as part of the framework. Chung et al. proposed a forensic framework to collect and analyze forensic data in a IoT eco-system~\cite{chung2017digital}. However, this solution is limited to Amazon Alexa ecosystem. 

\subsubsection{Smart Data Logging}
The idea of logging information from smart devices and apps is not novel. Several previous works have used this technique to have access to information as a result of smart app executions. In fact, the most popular IoT programming platforms provide the ways for logging smart app data \cite{wink, iris, casaverde, smartthingslogging}. Personalized open-source solutions can also be found \cite{simpleevent, thingspeak}. In this context, solutions like the proposed in \cite{fear} try to provide a comprehensive analysis of smart apps and smart devices logs for security purposes. In this work, the authors propose a platform-centric approach that looks at activities from smart apps for data provenance purposes. In this case, although this work is instrumental, the analysis is limited to consider the temporal relationship between devices and apps events without considering compromised devices, essential for forensic analysis.

\subsubsection{IoTDots and Information Flow Analysis in IoT}
Recently, information flow analysis (IFA) has moved from traditional research domains like android to IoT. For fairness, we believe it is necessary to evaluate the differences between IoTDots and some of the most popular IFA tools in the market today. In general, IoTDots achieves the capabilities of all the considered tools with high accuracy, low overhead, but while considering a comprehensive forensic model that includes challenges related to careless users, malicious users, malicious apps, and tampered devices. Additionally, more than policy enforcement or data provenance in IoT, IoTDots provides intelligence to match security policies in smart environments with interactions between smart apps, devices, and users. Table \ref{tab:IFA} summarizes the major differences between FlowFence \cite{fernandes2016flowfence}, ContextIoT \cite{jia2017contexiot}, ProvThings \cite{fear}, SaINT \cite{saint}, and IoTDots. For clarity, we used the analysis criteria similar to the one proposed in \cite{fear}. Then, we expanded the table features based on the IoTDots contributions. 

\textit{Compared to these prior works, IoTDots presents a comprehensive solution that automatically collects/stores forensically-relevant data from a smart environment considering (1) the interaction between users, devices, and apps inside the smart environment and (2) the security policies defined for every different environment. Additionally, our framework provides the means to analyze this data and detect anomalous activities from users and malicious behavior from users, smart apps, and devices. IoTDots flags those activities and behaviors that potentially violate the security policies of the smart environment and may help to hold the perpetrators accountable in a holistic manner during forensic investigations}.

\section{Conclusions} \label{sec:conclusion}
Smart devices and sensors deployed in smart environments have access to data that can be used for forensic purposes. Nonetheless, current smart app programming platforms do not provide any digital forensic capability to keep track of such information. Additionally, current forensic analysis solutions do not use information from smart apps and/or smart devices to perform forensic investigations. In this work, we introduce IoTDots, a novel framework used to extract forensically-relevant logs from smart apps and automatically analyze them later for forensic purposes. The framework has two main components: IoTDots-Modifier and IoTDots-Analyzer. The Modifier performs smart apps' source code analysis, detects forensically-relevant data points inside the smart app's source code, and inserts specific logs at compile time. Then, at runtime, the logs are sent to a remote IoTDots server. In a case of a forensic investigation, the Analyzer applies data processing and machine learning techniques to extract valuable and usable forensic information from the IoTDots logs. As per the evaluation results, IoTDots achieves over 98\% of accuracy on detecting user activities and over 96\% accuracy on detecting the behavior of users and smart apps in the smart environment. Additionally, IoTDots performs with very minimal to no overhead on the smart devices tested and very low overhead to the IoT cloud server resources. Finally, we have made IoTDots-Modifier available online.


\bibliographystyle{IEEEtran}
\bibliography{Bibtex}

\begin{thebibliography}{10}
\providecommand{\url}[1]{#1}
\csname url@samestyle\endcsname
\providecommand{\newblock}{\relax}
\providecommand{\bibinfo}[2]{#2}
\providecommand{\BIBentrySTDinterwordspacing}{\spaceskip=0pt\relax}
\providecommand{\BIBentryALTinterwordstretchfactor}{4}
\providecommand{\BIBentryALTinterwordspacing}{\spaceskip=\fontdimen2\font plus
\BIBentryALTinterwordstretchfactor\fontdimen3\font minus
  \fontdimen4\font\relax}
\providecommand{\BIBforeignlanguage}[2]{{%
\expandafter\ifx\csname l@#1\endcsname\relax
\typeout{** WARNING: IEEEtran.bst: No hyphenation pattern has been}%
\typeout{** loaded for the language `#1'. Using the pattern for}%
\typeout{** the default language instead.}%
\else
\language=\csname l@#1\endcsname
\fi
#2}}
\providecommand{\BIBdecl}{\relax}
\BIBdecl

\bibitem{fear}
Q.~Wang, W.~U. Hassan, A.~J. Bates, and C.~Gunter, ``Fear and logging in the
  internet of things,'' in \emph{Network and Distributed Systems Symposium
  (NDSS)}, Feb 2018.

\bibitem{icc}
L.~Babun, H.~Aksu, and A.~S. Uluagac, ``Identifying counterfeit smart grid
  devices: A lightweight system level framework,'' in \emph{2017 IEEE
  International Conference on Communications (ICC)}, May 2017, pp. 1--6.

\bibitem{thingspeak}
\BIBentryALTinterwordspacing
mattjfrank. (2018, May) Smartthings logging. [Online]. Available:
  \url{https://github.com/krlaframboise/SmartThings/blob/\\master/smartapps/krlaframboise/simple-event-logger.src/simple-event-logger.groovy}
\BIBentrySTDinterwordspacing

\bibitem{camera}
{Malware found in surveillance cameras sold through Amazon},
  \url{https://www.techworm.net/2016/04/malware-found-surveillance-cameras-sold-amazon.html},
  2017, [Online; accessed 10-January-2018].

\bibitem{hegarty2014digital}
R.~Hegarty, D.~J. Lamb, and A.~Attwood, ``Digital evidence challenges in the
  internet of things.'' in \emph{INC}, 2014, pp. 163--172.

\bibitem{phases}
{Y. Yusoff, R. Ismail and Z. Hassan}, ``Common phases of computer forensics
  investigation model,'' \emph{{International Journal of Computer Science \&
  Information Technology (IJCSIT)}}, vol.~3, no.~3, June 2011.

\bibitem{forensiciot}
\BIBentryALTinterwordspacing
R.~Udeshi. (2018) Why you need forensics in an iot world. [Online]. Available:
  \url{https://www.guidancesoftware.com/blog/digital-forensics/2017/10/03/why-you-need-forensics-in-an-iot-world}
\BIBentrySTDinterwordspacing

\bibitem{forensiciot2}
\BIBentryALTinterwordspacing
B.~P. Kondapally. (2018) What is iot forensics and how is it different from
  digital forensics? [Online]. Available:
  \url{https://securitycommunity.tcs.com/infosecsoapbox/articles/2018/02/27/what-iot-forensics-and-how-it-different-digital-forensics}
\BIBentrySTDinterwordspacing

\bibitem{magazine}
H.~Aksu, L.~Babun, M.~Conti, G.~Tolomei, and A.~S. Uluagac, ``Advertising in
  the iot era: Vision and challenges,'' \emph{IEEE Communications Magazine},
  pp. 1--7, 2018.

\bibitem{variety}
\BIBentryALTinterwordspacing
U.~Salama. (2018) Smart forensics for the internet of things (iot). [Online].
  Available:
  \url{https://securityintelligence.com/smart-forensics-for-the-internet-of-things-iot/}
\BIBentrySTDinterwordspacing

\bibitem{Official}
\BIBentryALTinterwordspacing
Samsung. (2018) Smartthings official app repository. [Online]. Available:
  \url{https://github.com/SmartThingsCommunity}
\BIBentrySTDinterwordspacing

\bibitem{saint}
\BIBentryALTinterwordspacing
Z.~B. Celik, L.~Babun, A.~K. Sikder, H.~Aksu, G.~Tan, P.~McDaniel, and A.~S.
  Uluagac, ``Sensitive information tracking in commodity iot,'' in \emph{27th
  {USENIX} Security Symposium ({USENIX} Security 18)}.\hskip 1em plus 0.5em
  minus 0.4em\relax Baltimore, MD: {USENIX} Association, 2018. [Online].
  Available:
  \url{https://www.usenix.org/conference/usenixsecurity18/presentation/celik}
\BIBentrySTDinterwordspacing

\bibitem{SaINTWeb}
\BIBentryALTinterwordspacing
{L. Babun, Z. Berkay Celik and A. Kumar Sikder}. (2018) Saint project.
  [Online]. Available: \url{http://saint-project.appspot.com/}
\BIBentrySTDinterwordspacing

\bibitem{Community}
\BIBentryALTinterwordspacing
Samsung. (2018) Smartthings community forum for third-party apps. [Online].
  Available: \url{https://community.smartthings.com/}
\BIBentrySTDinterwordspacing

\bibitem{openHABMarket}
\BIBentryALTinterwordspacing
OpenHAB. (2018) Openhab iot app market (eclipse market place). [Online].
  Available: \url{http://docs.openhab.org/eclipseiotmarket}
\BIBentrySTDinterwordspacing

\bibitem{appleMarket}
\BIBentryALTinterwordspacing
Apple. (2018) Apple's homekit app market. [Online]. Available:
  \url{https://support.apple.com/en-us/HT204893}
\BIBentrySTDinterwordspacing

\bibitem{smartThings-review}
\BIBentryALTinterwordspacing
Samsung. (2018) Smartthings code review guidelines and best practices.
  [Online]. Available:
  \url{http://docs.smartthings.com/en/latest/code-review-guidelines.html}
\BIBentrySTDinterwordspacing

\bibitem{OpenHabGuideline}
\BIBentryALTinterwordspacing
OpenHAB. (2018) Openhab iot app submission guideline. [Online]. Available:
  \url{https://marketplace.eclipse.org/content/eclipse-marketplace-publishing-guidelines}
\BIBentrySTDinterwordspacing

\bibitem{AppleHomekitReview}
\BIBentryALTinterwordspacing
Apple. (2018) Apple's homekit submission guideline. [Online]. Available:
  \url{https://developer.apple.com/app-store/review/guidelines}
\BIBentrySTDinterwordspacing

\bibitem{SmartThingsAPI}
\BIBentryALTinterwordspacing
Samsung. (2018) Smartthings official api documentation. [Online]. Available:
  \url{http://docs.smartthings.com/en/latest/ref-docs/reference.html}
\BIBentrySTDinterwordspacing

\bibitem{OpenHABPrivacy}
\BIBentryALTinterwordspacing
OpenHAB. (2018) Openhab privacy statement. [Online]. Available:
  \url{http://www.myopenhab.org/privacy}
\BIBentrySTDinterwordspacing

\bibitem{GroovyVisitor}
\BIBentryALTinterwordspacing
GroovyCodeVisitor. (2018) Groovycodevisitor: An implementation of the groovy
  visitor patterns. [Online]. Available: \url{http://docs.groovy-lang.org/docs}
\BIBentrySTDinterwordspacing

\bibitem{smartThings-documentation}
\BIBentryALTinterwordspacing
Samsung. (2018) Smartthings official developer documentation. [Online].
  Available: \url{http://docs.smartthings.com}
\BIBentrySTDinterwordspacing

\bibitem{6thSense}
A.~K. Sikder, H.~Aksu, and A.~S. Uluagac, ``{6thSense: A Context-aware
  Sensor-based Attack Detector for Smart Devices},'' in \emph{USENIX Security},
  2017.

\bibitem{samsung}
\BIBentryALTinterwordspacing
Samsung. (2018) Samsung smartthings. [Online]. Available:
  \url{https://www.smartthings.com/}
\BIBentrySTDinterwordspacing

\bibitem{iotbench}
\BIBentryALTinterwordspacing
{L. Babun, Z. Berkay Celik and A. Kumar Sikder}. (2018) Iotbench repository.
  [Online]. Available: \url{https://github.com/IoTBench}
\BIBentrySTDinterwordspacing

\bibitem{ContexIoT-repo}
{ContexIoT attacks for SmartThings programs using existing adversary
  techniques}, \url{https://sites.google.com/site/iotcontextualintegrity/home},
  2017, [Online; accessed 09-January-2018].

\bibitem{fernandes2016security}
E.~Fernandes, J.~Jung, and A.~Prakash, ``{Security Analysis of Emerging Smart
  Home Applications},'' in \emph{IEEE Security and Privacy (SP)}, 2016.

\bibitem{smartTV-hack}
\BIBentryALTinterwordspacing
J.~Zhu. (2018) Android-based smart tvs hit by backdoor spread via malicious
  app. [Online]. Available:
  \url{https://blog.trendmicro.com/trendlabs-security-intelligence/android-based-smart-tvs-hit-by-backdoor-spread-via-malicious-app/}
\BIBentrySTDinterwordspacing

\bibitem{homeseer}
\BIBentryALTinterwordspacing
HomeSeer. (2018) Homeseer. [Online]. Available: \url{https://homeseer.com/}
\BIBentrySTDinterwordspacing

\bibitem{homeos}
\BIBentryALTinterwordspacing
Microsoft. (2018) Homeos. [Online]. Available:
  \url{https://www.microsoft.com/en-us/research/project/homeos-enabling-smarter-homes-for-everyone/}
\BIBentrySTDinterwordspacing

\bibitem{chung2012digital}
H.~Chung, J.~Park, S.~Lee, and C.~Kang, ``Digital forensic investigation of
  cloud storage services,'' \emph{Digital investigation}, vol.~9, no.~2, pp.
  81--95, 2012.

\bibitem{oriwoh2013internet}
E.~Oriwoh, D.~Jazani, G.~Epiphaniou, and P.~Sant, ``Internet of things
  forensics: Challenges and approaches,'' in \emph{Collaborative Computing:
  Networking, Applications and Worksharing (Collaboratecom), 2013 9th
  International Conference Conference on}.\hskip 1em plus 0.5em minus
  0.4em\relax IEEE, 2013, pp. 608--615.

\bibitem{watson2016digital}
S.~Watson and A.~Dehghantanha, ``Digital forensics: the missing piece of the
  internet of things promise,'' \emph{Computer Fraud \& Security}, vol. 2016,
  no.~6, pp. 5--8, 2016.

\bibitem{kebande2016generic}
V.~R. Kebande and I.~Ray, ``A generic digital forensic investigation framework
  for internet of things (iot),'' in \emph{Future Internet of Things and Cloud
  (FiCloud), 2016 IEEE 4th International Conference on}.\hskip 1em plus 0.5em
  minus 0.4em\relax IEEE, 2016, pp. 356--362.

\bibitem{zawoad2015faiot}
S.~Zawoad and R.~Hasan, ``Faiot: Towards building a forensics aware eco system
  for the internet of things,'' in \emph{Services Computing (SCC), 2015 IEEE
  International Conference on}.\hskip 1em plus 0.5em minus 0.4em\relax IEEE,
  2015, pp. 279--284.

\bibitem{chung2017digital}
H.~Chung, J.~Park, and S.~Lee, ``Digital forensic approaches for amazon alexa
  ecosystem,'' \emph{Digital Investigation}, vol.~22, pp. S15--S25, 2017.

\bibitem{wink}
\BIBentryALTinterwordspacing
Wink. (2018, May) Wink. [Online]. Available: \url{https://www.wink.com/}
\BIBentrySTDinterwordspacing

\bibitem{iris}
\BIBentryALTinterwordspacing
Lowe's. (2018, May) Iris by lowe's. [Online]. Available:
  \url{https://www.irisbylowes.com/}
\BIBentrySTDinterwordspacing

\bibitem{casaverde}
\BIBentryALTinterwordspacing
MiCasaVerde. (2018, May) Logs - micasaverde. [Online]. Available:
  \url{http://wiki.micasaverde.com/index.php/Logs}
\BIBentrySTDinterwordspacing

\bibitem{smartthingslogging}
\BIBentryALTinterwordspacing
Samsung. (2018, May) Smartthings official logging. [Online]. Available:
  \url{http://docs.smartthings.com/en/latest/tools-and-ide/logging.html}
\BIBentrySTDinterwordspacing

\bibitem{simpleevent}
\BIBentryALTinterwordspacing
K.~LaFramboise. (2018, May) Simple event logger. [Online]. Available:
  \url{https://github.com/krlaframboise/SmartThings/blob/\\master/smartapps/krlaframboise/simple-event-logger.src/simple-event-logger.groovy}
\BIBentrySTDinterwordspacing

\bibitem{fernandes2016flowfence}
E.~Fernandes, J.~Paupore, A.~Rahmati, D.~Simionato, M.~Conti, and A.~Prakash,
  ``{FlowFence: Practical Data Protection for Emerging IoT Application
  Frameworks},'' in \emph{USENIX Security}, 2016.

\bibitem{jia2017contexiot}
Y.~J. Jia, Q.~A. Chen, S.~Wang, A.~Rahmati, E.~Fernandes, Z.~M. Mao,
  A.~Prakash, and S.~J. Unviersity, ``{ContexIoT: Towards Providing Contextual
  Integrity to Appified IoT Platforms},'' in \emph{NDSS}, 2017.

\end{thebibliography}

\end{document}